\documentclass[journal,10pt]{IEEEtran}
\usepackage{cite}      
\usepackage{graphicx}  
\usepackage{url}
\usepackage{balance}
\usepackage{times}
\usepackage{amsmath}   
\usepackage{pdflscape}
\usepackage[normalem]{ulem}
\usepackage{mathtools}
\usepackage{nicefrac}
\usepackage[none]{hyphenat}
\usepackage{array}
\usepackage{fontenc}
\usepackage{color}
\usepackage{etoolbox}
\usepackage{bm}
\usepackage{comment}

\usepackage{amsmath,amsthm,amssymb, amsfonts}
\usepackage{mathtools}
\usepackage{wrapfig}
\usepackage{nomencl}
\usepackage{algorithm, algorithmic}
\usepackage{accents}

\newcommand{\argmax}{\arg\!\max}

\newlength{\dhatheight}

\newcolumntype{P}[1]{>{\centering\arraybackslash}p{#1}}
\newcolumntype{M}[1]{>{\centering\arraybackslash}m{#1}}
\usepackage{verbatim} 
\usepackage{float}
\usepackage{placeins}
\usepackage{comment}
\usepackage{pifont}
\newcommand{\xmark}{\ding{55}}%

\DeclareMathOperator{\Ga}{\mathbf{g}_p}

\DeclareMathOperator{\stx}{\mathbf{s}_{tx}}
\DeclareMathOperator{\srx}{\mathbf{s}_{rx}^k}
\DeclareMathOperator{\K}{\mathbb{\mathbf{K}}}
\DeclareMathOperator{\tilK}{\mathbb{\mathbf{\tilde{K}}}}
\DeclareMathOperator{\snr}{S_{k}}
\DeclareMathOperator{\snri}{S_{\overline{k}}}

\usepackage{xparse}

\begin{document}

\setlength{\textfloatsep}{0pt}
\setlength{\intextsep}{10pt plus 2pt minus 2pt}

\title{Beam Alignment in Multipath Environments for Integrated Sensing and Communication using Bandit Learning}

\author{
	\IEEEauthorblockN{Akanksha~Sneh,  Shobha~Sundar~Ram, \textit{Senior Member, IEEE}, Sumit~J~Darak, \textit{Senior Member, IEEE} and Aakanksha Tewari }
\thanks{All authors are with ECE Dept., IIIT-Delhi, India. E-mail: \{akankshas, shobha, summit,aakankshat\}@iiitd.ac.in. This work is supported by the funding received from Meity CC \& BT, IIT-H TiHAN and Qualcomm Innovation Fellowship 2022 India.} }

\maketitle

\begin{abstract}
Prior works have explored multi-armed bandit (MAB) algorithms for the selection of optimal beams for millimeter-wave (mmW) communications between base station and mobile users. However, when the number of beams is large, the existing MAB algorithms are characterized by long exploration times, resulting in poor overall communication throughput. In this work, we propose augmenting the upper confidence bound (UCB) based MAB with integrated sensing and communication (ISAC) to address this limitation. The premise of the work is that the radar and communication functionalities share the same field-of-view and that communication mobile users are detected by the radar as mobile targets. The radar information is used for significantly reducing the number of candidate beams for the UCB, resulting in an overall reduction in the exploration time. Further, the radar information is used to estimate the realignment time in quasi-stationary scenarios. We have realized the MAB and radar signal processing algorithms on the system on chip (SoC) via hardware-software co-design (HSCD) and fixed-point analysis. We demonstrate the significant gain in execution time using accelerators. The simulations consider complex propagation channels involving direct and multipath, with simple and extended radar targets in the presence of significant static clutter.
The resulting experiments show that the proposed ISAC-based MAB achieves a 35\% reduction in the overall exploration time and 1.4 factor higher throughput as compared to the conventional MAB that is based only on communications.   
\end{abstract}


\section{Introduction}
\label{sec: Intro}
Next-generation (5G/6G) communications are expected to support peak data rates between 1 and 20 Gigabits per second, at very high connection densities in urban scenarios (of the order of a million devices per square area) with above 99\% reliability and latency below 10 ms \cite{dang2020should}. These specifications will enable diverse applications such as ubiquitous virtual reality, augmented reality, connected cars, indoor positioning systems, and industrial robotics \cite{patzold20195g}. The delivery of these services at the desired performance metrics is predicated on suitably handling spectrum congestion issues \cite{luong2021radio}. 

Integrated sensing and communication (ISAC) has been identified as a key instrument towards the realization of these objectives by supporting dual functional utilization of the spectrum \cite{liu_joint_2020}. Research on ISAC must be distinguished from the other works that have focused on spectrum sharing between radar and communications. Some of the earliest works considered the interference between coexisting radar and communication systems \cite{deng2013interference}. Examples of these systems are the shared spectrum between 4G-LTE and early warning airborne radar systems; IEEE 802.11a/h/j WLAN networks and vessel traffic service radars \cite{liu_joint_2020}. The second body of works has explored the use of communication transmitters - navigation satellites, WiFi, and radio - as opportunistic illuminators for passive radar receivers \cite{zeng2016wireless}. {\color{black}Further, in \cite{wang2023towards}, authors proposed a framework to employ the MAC layer provisions of the IEEE 802.11ad protocol for indoor radar sensing to support various IoT applications. } \textcolor{black}{In ISAC, sensing and communication functionalities are realized with shared waveform and hardware and accomplished through various multiplexing techniques. Frequency division multiplexing-based ISAC, such as those described in \cite{chiriyath2017radar,bicua2018radar}, involves allocations of sensing and communication functionalities to different subcarriers of the OFDM waveform while guaranteeing both target localization information and achievable data rate constraints for radar sensing and communications, respectively. Spatial division multiplexing-based ISAC is majorly employed in multiple-input multiple-output (MIMO) and massive MIMO technologies \cite{liu2018mu}. In this technique, sensing over orthogonal spatial resources, such as different antenna groups, is used. For line-of-sight scenarios, sensing and communication waveforms can be transmitted over different spatial beams \cite{d2019communications}. However, if the communication channel model constitutes several scattering paths, the sensing waveform may be projected into its null space to avoid interfering with the communication functionality \cite{mahal2017spectral}. Code division multiplexing-based ISAC involves sensing and communication signals carried by orthogonal/quasi-orthogonal sequences \cite{ye2022low}. Lastly, time-division multiplexing (TDM) based ISAC \cite{han2013joint} is the most adopted technique as it can be conveniently implemented into the existing commercial systems. In TDM ISAC, the duration of one cycle is split between the radar and communication functionalities. Over the recent years, TDM-based ISAC has been studied for several commercial wireless standards, such as IEEE 802.11p and IEEE 802.11ad, where the channel estimation field (CEF)/pilot signals, originally designed for channel estimation, are exploited for radar sensing \cite{kumari_ieee_2018,grossi2018opportunistic,duggal2019micro,sneh2023ieee}. }  Further, the strategies may adopt embedding radar signals with information \cite{hassanien2019dual,wu2020waveform} or using communication signals for radar sensing purposes. For example, a radar-based collision detection system was implemented using IEEE 802.11a/g \cite{daniels2017forward}. At millimeter-wave (mmW) frequencies, IEEE 802.11ad radar has been proposed and developed for automotive target detection and tracking \cite{kumari_ieee_2018,duggal2020doppler}. More recently, orthogonal time-frequency space signals that produce delay-Doppler ambiguity diagrams have also been explored for ISAC \cite{shi2023integrated}.

Unlicensed spectrum at mmW frequencies offers high bandwidth (above 2GHz) that can potentially support Gigabits per second data rates for supporting vehicular communications \cite{choi2016millimeter}. The state-of-the-art vehicular communication protocols at sub-6GHz such as dedicated short-range communications \cite{kenney2011dedicated}, cellular LTE \cite{wang2018cellular} or device to device LTE \cite{asadi2014survey} support much lower data rates and have very high latency. 
However, unlike microwave frequencies, mmW bands are characterized by significant atmospheric absorption necessitating narrow directional beams. Further, high-mobility vehicular environments are challenged by significant Doppler shifts due to the high carrier frequencies, and significant multipath and clutter. All of these factors necessitate rapid beam alignment of the narrow beams at both the base station (BS) and mobile user (MU). The focus of this work is to facilitate rapid beam alignment in ISAC systems at the BS in multipath environments through online learning.

Among the many different classes of online learning algorithms, multi-armed bandit (MAB) algorithms from the reinforcement learning framework have emerged as valuable tools for optimal beam selection at BS usually with the acknowledgment feedback from MU as reward \cite{nasim2020learning,aykin2020mamba}.  Optimal codebook and exponential weights algorithm were studied in \cite{chafaa2019adversarial} for beam alignment and tracking in dynamic mmW channels. More recently MAB algorithms have been explored for cognitive radar/radio environments \cite{9114698,9322256,9455255,9527128}. In \cite{9455255,9527128}, the authors exploited stochastic and adversarial MAB algorithms for waveform selection in pulse-agile radar systems for optimizing the detection performance while mitigating harmful sidelobe levels which enhances its robustness to jamming attacks. In \cite{9114698,9322256}, contextual bandit learning algorithms and other variants of reinforcement learning were examined for adapting radar waveforms for managing interference during spectrum sharing with a non-cooperative cellular network. 

In this work, we consider the radar and communication systems to share the same spatial field of view. Among the multiple radar targets, in this field of view, are communication MU along with clutter and multipath that give rise to both missed detections and false alarms. Our work considerably expands upon our preliminary work, where we proposed a simple radar-enhanced MAB algorithm to reduce the beam search times using the estimates of the target's amplitude and Doppler \cite{sneh2023radar}. The prior work was premised on key assumptions regarding stationary environmental conditions for single simple point-target scenarios. Further, important considerations regarding the real-time performance in realistic environments with multipath and non-idealistic radar operating curves were omitted, which are considered here. In this work, the radar estimates of each MU's range and Doppler are used in conjunction with the UCB algorithm to reduce the number of candidate beams that must be scanned before eventually identifying the optimal beams for communication with MU. Second, the proposed algorithm is envisaged for \emph{quasi-stationary} environmental conditions where it forecasts when we must recommence scanning for new optimal beams. The performance of the algorithm is evaluated for mobile road users such as cars, and pedestrians modeled as extended radar targets with multiple-point scatterers. These targets undertake trajectories in realistic electromagnetic propagation environments characterized by static clutter and multipath.  Third, the proposed radar signal processing (RSP) and MAB algorithms are realized on an edge system-on-chip (SoC) platform comprising of multi-core processor and field programmable gate array (FPGA) platforms via hardware-software co-design and fixed-point analysis. {\color{black} Using the hardware timing results, we demonstrate that the proposed approach offers 1.4 factor improvement in throughput over the state-of-the-art.} 

The paper is organized as follows. \textcolor{black}{We present a literature study of relevant works in Section \ref{sec:lit_review}.} Next, in Section \ref{sec:SignalModel}, we present the signal model for the transmitter and the receiver followed by the proposed methodology in Section \ref{sec:algos}.  Then, we discuss the simulation experiments in Section \ref{sec:Simulationsetup} and present the results in Section \ref{sec:Results}. We discuss the computational resource utilization and timing analysis of the hardware-software codesign implementation of the algorithms in Section VII. We conclude the paper with a summary of the findings and a discussion on future works in Section \ref{sec:Conclusion}.

\emph{Notation:} Scalar variables, vectors and matrices are indicated with $v$, $\mathbf{v}$ and $\mathbf{V}$ respectively. \textcolor{black}{Vector superscripts $T$ and $H$ denote transpose and Hermitian transpose respectively.  The symbols $\ast$ and $\odot$ denote the convolution and element-wise multiplication operations.} 
We use the square braces, $[\cdot]$, to indicate discrete-time sequences, and the curly braces, $(\cdot)$, to indicate continuous time signals.
{\color{black}
\section{Relevant Works}
\label{sec:lit_review}
In this section, we present a comprehensive literature study of beam alignment in mmW communications. The prior art is broadly classified into two categories: beam alignment in ISAC and bandit learning-based beam alignment in mmW communications. These have been elaborated as follows:

\subsubsection{Beam Alignment in ISAC}  The use of an auxiliary radar sensor for detecting mobile users was considered in \cite{va_beam_2015}. However, this method increases the cost and complexity of the system. Further, the radar and communication functionalities would have to be synchronized as well as managed for interference. ISAC systems have been identified as a key instrument for beam alignment in mmW communication, such as those proposed by \cite{gonzalez2016radar,zhang2018multibeam,muns2019beam,liu2020radar,chang2021joint,sneh2023ieee}. Here no separate hardware/spectrum/synchronization or interference management is required to support both functionalities.
The authors in \cite{gonzalez2016radar} presented radar-aided target detection and estimation of spatial correlation using a compresssive approach to enable beam alignment for vehicle-to-infrastructure scenarios. However, this work is suited for line-of-sight channels, which may not be realistic in more complex scenarios. The work presented in \cite{zhang2018multibeam} proposed one-dimensional compressive sensing-based estimations of the angle, delay, and Doppler frequency of the multipath signals. The authors in \cite{muns2019beam} used radar signal processing (RSP) for target localization and beamforming. However, the ISAC transceiver must be operated in full-duplex mode in order to detect short-range targets, and RSP needs to be carried out for every packet, leading to high latency and computational complexity. Predictive beamforming based on Kalman filtering along with RSP-based target detection was proposed in \cite{liu2020radar}. The work presented in \cite{chang2021joint} proposed a joint communication and control algorithm where both the transmission data rate requirement in communications and event-triggered control policy are considered in the beam alignment design. Again, there is an increase in complexity due to the control mechanism. In our previous work \cite{sneh2023ieee}, we presented an end-to-end software prototype of an IEEE 802.11ad-based joint radar communication transceiver, which enables RSP-based beamforming. Here, the azimuth angles of the targets were determined through the digital beamforming at the base station receiver. This leads to hardware complexity due to multiple RF chain blocks required for each element of the receiver antenna array. Note that in none of the above works, reinforcement learning is used for the beam alignment.

\subsubsection{Bandit Learning Based Beam Alignment} There have been several recent works that have researched multi-armed bandit algorithms for reducing the exploration time of the best beams, in analog beamforming scenarios\cite{zhang2020beam,booth2019multi,nasim2020learning,cheng2019fast,aykin2020mamba}. In \cite{booth2019multi}, authors utilized the linear Thompson sampling method for beamforming and Kalman filter for tracking, but this resulted in increased latency and performance degradation for a large number of beams and static clutters. A bandit-inspired beam searching
scheme to reduce the number of measurements in high-speed trains was proposed in \cite{cheng2019fast}. Further, in \cite{zhang2020beam}, authors presented an integrated greedy policy and upper confidence bound strategy for the beam alignment. However, they assumed a certain structure on beam distribution and used it for faster beam selection via MAB. Note that we do not make such an assumption. The authors in \cite{nasim2020learning} presented an adaptive combinatorial Thompson sampling algorithm for the appropriate selection of a set of beams for multiple users in a high-mobility vehicular environment. Here, the complexity increased with the number of candidate beams \cite{santosh2022multiarmed}. \cite{aykin2020mamba} developed a reinforcement learning algorithm called adaptive Thompson sampling (ATS) that embodied the selection of appropriate beams
and transmission rates along these beams. ATS uses prior beam quality information collected through the initial access and updates it whenever acknowledgment feedback is obtained from the user. They further adopted MCS to improve the throughput. However, performance degradation happened with a large number of beams. For beam tracking, additional feedback using initial access was required.

The challenges and limitations discussed in both radar-enabled beam alignment of bandit learning-based beam alignment in mmW communications can be overcome with our proposed approach. 
The unique contribution of our work is to use the ISAC model with multi-armed bandit algorithms for rapid beam alignment in analog beamforming scenarios for mmW communications. Radar detection-based decision-making will be much faster than communication metric-based decision-making due to multiple reasons. First, radar signal processing can commence immediately after a scattered signal is received from the mobile target, which is nearly instantaneous, unlike communication signal processing, which can be initiated only after receiving an acknowledgment through uplink communication. Second, the exploration time is substantially reduced by restricting the number of candidate beams that must be scanned. Due to these factors, the overall exploration time will be reduced, resulting in rapid beam alignment and improved overall communication throughput. For a quasi-stationary environment, the proposed radar sensing-based change detection is compute and memory-efficient and does not need prior knowledge of the number of change points compared to existing approaches. Further, hardware implementation on system-on-chip (SoC) via hardware-software co-design and word length analysis may provide the research community with a better understanding of the 802.11ad-based ISAC-MAB framework. The concept of radar-enhanced MAB may also be expanded to other ISAC models as well and is not limited to the IEEE 802.11ad signal model. {\color{black} For easier understanding, we compare the key novelties of the proposed architecture in our work with respect to the prior art regarding beam alignment in ISAC for mmW communications in Table \ref{tab:literature_review}.}
\begin{table*}[htbp]
{\color{black}
 \centering
  \caption{{\footnotesize \color{black}Literature survey comparing the contributions of the proposed work with the prior art}}
  \renewcommand{\arraystretch}{1}
 \resizebox{\textwidth}{!}{
  \begin{tabular}{M{2cm}|M{2cm}|M{2cm}|M{2cm}|M{2cm}|M{2cm}|M{2cm}}\hline
       Reference & ISAC Transceiver Codesign & RSP and Bandit Learning Based Beam Selection &RSP on Edge  & Extended Targets Tracking & Hardware- Software Codesign & RSP Based Change Detection \\\hline\hline

  \cite{kumari2017ieee}&\checkmark &\xmark& \xmark &\checkmark & \xmark & \xmark
\\\hline
\cite{liu2018toward}&\checkmark&\xmark&\xmark &\xmark&\xmark & \xmark \\\hline
  \cite{hassanien2019dual} &\checkmark &\xmark&\xmark& \xmark  & \xmark & \xmark
\\\hline
 
\cite{muns2019beam}&\checkmark&\xmark & \xmark&\xmark  &\xmark & \xmark
\\\hline
  \cite{duggal2020doppler}& \checkmark  &\xmark& \xmark &\checkmark &\xmark & \xmark
\\\hline
       
   \cite{liu2020joint}&\checkmark&\xmark & \xmark&\xmark  &\xmark & \xmark
\\\hline

 \cite{liu2020radar}& \checkmark&\xmark & \xmark&\xmark  &\xmark & \xmark 
\\\hline   

 \cite{kumari2021jcr70}& \checkmark& \xmark & \checkmark & \checkmark  &\xmark  & \xmark
\\\hline
 \cite{pegoraro2021rapid}& \checkmark&\xmark & \checkmark&\checkmark &\xmark  & \xmark
\\\hline
  
\cite{sneh2023radar}& \checkmark&\checkmark & \xmark&\xmark  &\xmark  & \xmark
\\\hline
Proposed work  &\checkmark &\checkmark&\checkmark &\checkmark&\checkmark &\checkmark 
\\\hline
   \end{tabular}}
   \vspace{-0.2cm}
   \label{tab:literature_review}}
\end{table*}

\section{ISAC Signal Model and  Signal Processing}
\label{sec:SignalModel}
In this work, we consider a BS with ISAC capability that supports analog beamforming at both the transmitter (TX) and receiver (RX). We assume that the radar and communication functionalities of the ISAC share the same beamforming hardware and spatial field of view. In other words, the radar also uses analog beamforming hardware to examine the same set of beams as the communications. Therefore, the MU for communications are mobile targets from the radar's point of view and the two functionalities share the resources through time division multiplexing 
{\color{black} as shown in Fig.\ref{fig:JRCTiming}. 
\begin{figure}[htbp]
\centering
\includegraphics[scale=0.52]{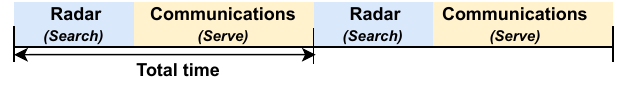}
\caption{\footnotesize \textcolor{black}{Duty cycle of radar search time over total time due to shared functionality with communications.}}
\label{fig:JRCTiming}
\end{figure}
Note that the radar search time is distinct from the exploration time in the MAB framework. The longer the radar search time. Hence, for a fixed total time interval, the \emph{longer} the radar takes to search through analog beamforming of the beams in the JRC field of view to detect and localize the mobile users, the \emph{shorter} is the communication service time available for those mobile users, and lower resultant throughput. The choice of the radar duty cycle (duration of radar functionality over the total time interval for radar and communication) trades off between communication throughput and the radar detection performance. The analysis governing the optimum choice of the radar duty cycle (including radar packet times, the pulse repetition interval (PRI), and the processing times) must factor in both the radar detection metrics and the subsequent communication throughput. This optimization theory is discussed in length in \cite{ram2022optimization}. }
The ISAC transmit waveform is scattered back by the targets and reaches the RX at the BS. Here, it is processed to estimate the target's properties (range and Doppler). The radar information is used to subsequently direct the beams during the data communication phase. The BS must detect and locate single or multiple MU in the field-of-view (FoV) in the presence of clutter and multipath by exploring multiple beams, one at a time.  

\subsection{Signal Model}
In this section, we present signal models for the ISAC transmitter and receiver. {\color{black} The literature reports different waveforms employed to support ISAC, which can be broadly classified into two categories - sensing-centric waveforms and communication-centric waveforms. The sensing-centric waveforms aim to incorporate the communication functionality into the existing sensing framework where the sensing performance is the primary goal \cite{roberton2003integrated,slavik2019cognitive}. On the other hand, the communication-centric waveform enables the sensing and localization of the target along with the communication \cite{johnston2022mimo,keskin2021mimo}.  Recently, standardized mmW communication protocols have been used in literature \cite{wang2023towards,zhang2021design,grossi2018opportunistic,kumari_ieee_2018,duggal2020doppler,sneh2023ieee} that are found to be suitable for the ISAC framework. In our work, the 512-Golay sequence $\mathbf{g}_{u_{512}}$ (shown in pink color) in the channel estimation (CE) of the preamble of IEEE 802.11ad PHY packet in 
Fig.~\ref{fig:phy_packet}.
    \begin{figure}[htbp]
    \centering
    \includegraphics[scale=0.54]{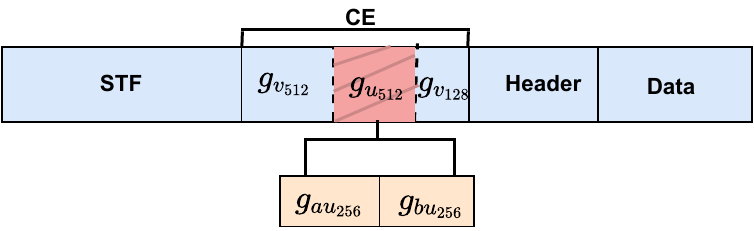}
    \caption{\footnotesize{\color{black}IEEE 802.11ad PHY frame with radar waveform embedded in it.}}
    \label{fig:phy_packet}   
\end{figure}
is extracted out and separately sent as a radar waveform.
The Golay sequences are noted for their perfect autocorrelation properties, enabling radar range estimation of the targets.} \textcolor{black}{However, the perfect autocorrelation property is disturbed when a moving target induces a Doppler phase shift.} This issue can be resolved by choosing appropriate Doppler resilient Golay sequences in consecutive packets denoted by $\Ga$. These are sampled at a frequency of $F_s=1/T_s$ and converted to analog using a digital-to-analog converter.  
The analog radar transmitted signal, $\stx$ in terms of fast time $\tau$ for every $p^{th}$ packet is given by
\par\noindent\small
\begin{align}
\stx(\tau,p) = \left(\sum_{n=0}^{N-1}\Ga[n]\delta(\tau-nT_s-pT_p) \right).
\end{align}
\normalsize
Here $T_p$ is the duration of one pulse repetition interval (PRI) and $\delta(\cdot)$ is the Dirac-delta function. The signal consisting of $P$ such pulses is upconverted to the carrier frequency, amplified, and provided to a uniform linear array of $Q$ antennas with half-wavelength spacing that are weighted with $\mathbf{w}_{\phi_k} \in \mathbb{C}^{Q \times 1}$ to direct the transmission along a pre-determined direction $\phi_k$.

We assume that the signal is incident upon a $b^{th}$ MU at a range $r_b$, azimuth $\phi_b$, moving with a radial velocity component $v_b$.  \textcolor{black}{At the receiver, $\mathbf{w}_{\phi_k}^T \in \mathbb{C}^{1 \times Q}$ correspond to the weights assigned to enable analog beamforming along a pre-determined direction $\phi_k$.} The scattered signal at the $k^{th}$ beam of the receiver array, is from a collection of $B$ MU, $M$ multipath components (MC) and $L$ static clutter scatterers (SCS). This signal is given in terms of fast time, $\tau$ (across a single packet) and slow time, $t$, across multiple packets as
\par\noindent\small
\begin{align}
\label{eq:LOScond}
\nonumber\srx(\tau,t) = \sum_{b=1}^B \sigma_{b}\mathbf{w}_{\phi_k}^T\mathbf{u}_{\phi_b}\mathbf{u}_{\phi_b}^T\mathbf{w}_{\phi_k}\stx(\tau-\tau_{b}-t)e^{j2\pi f_bt}+\\\nonumber\sum_{m=1}^M\sum_{b=1}^B \mathbf{w}_{\phi_k}^T\mathbf{H}_m\mathbf{w}_{\phi_k}\stx(\tau-\tau_{m}-t)e^{j2\pi f_{m}t} \\
+\sum_{l=1}^L \sigma_{l}\mathbf{w}_{\phi_k}^T\mathbf{u}_{\phi_l}\mathbf{u}_{\phi_l}^T\mathbf{w}_{\phi_k}\stx(\tau-\tau_{l}-t)+
\zeta. 
\end{align}
\normalsize
Here, $\tau_b = \frac{2r_b}{c}$, and $f_b = \frac{2v_bf_c}{c}$ are the range-induced delay and Doppler frequency shift from the MU respectively. Further, $\sigma_b$ incorporates the scattering coefficient, path loss factor, the transmitting power, system loss factors, and each elemental antenna's gain, and $\zeta$ represents the complex white Gaussian noise at the radar receiver. MmW channels are complex propagation environments characterized by both direct and multipath components. In the above expression, the direct path to the targets is modeled in the first term using the steering vector, $\mathbf{u}_{\phi_b} = [1\;e^{j\pi\sin\phi_b} \cdots e^{j(Q-1)\pi\sin\phi_b}]^T$, from the radar to the target based on the assumption that the elements are spaced half-wavelength apart. The second term in the expression represents the multipath components. Here, the delay ($\tau_m$), azimuth ($\phi_m$) and Doppler shift ($f_m$) of the multipath components are drawn from uniform distributions from the radar FoV and  
 \par\noindent\small
\begin{align}
{\color{black}
\mathbf{H}_m = \alpha_m \odot \mathbf{u}_{\phi_m}\mathbf{u}_{\phi_b}^T, }
\end{align}
\normalsize
{\color{black}models the multipath propagation from the transmitting to receiving antennas. Further, $\alpha_m$ models the complex channel gain corresponding to the $m^{th}$ multipath component, and $\mathbf{u}_{\phi_m}$ is the steering vector along $\phi_m$ direction. }
In \eqref{eq:LOScond}, we also have static clutter scatterers (SCS) that are detected by the radar but are distinct from the MU. These are characterized by RCS $\sigma_l$ and delay $\tau_l$ and azimuth $\phi_l$.
\begin{figure*}[htbp]
    \centering
    \includegraphics[scale=0.48]{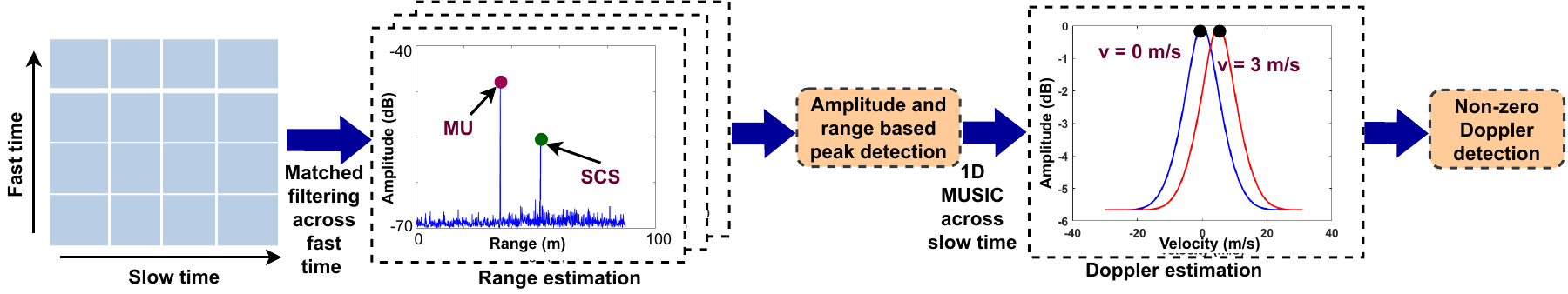}
    \caption{\footnotesize Block diagram of ISAC-MAB algorithm.}
    \label{fig: block_diagram}
\end{figure*}

\subsection{Radar signal processing} 
The received signal is downconverted, digitized, and organized as a radar data square as shown in Fig.~\ref{fig: block_diagram} with fast-time samples (dimension $n$) across multiple slow-time packets (dimension $p$). 
\par\noindent\small
\begin{align}
\nonumber\srx[n,p] = \sum_{b=1}^B \sigma_{b}\mathbf{w}_{\phi_k}^T\mathbf{u}_{\phi_b}\mathbf{u}_{\phi_b}^T\mathbf{w}_{\phi_k}\stx(nT_s-\tau_{b}-pT_p)e^{j2\pi f_bpT_p}+\\\nonumber\sum_{m=1}^M\sum_{b=1}^B \mathbf{w}_{\phi_k}^T\mathbf{H}_m\mathbf{w}_{\phi_k}\stx(nT_s-\tau_{m}-pT_p)e^{j2\pi f_{m}pT_p} \\
+\sum_{l=1}^L \sigma_{l}\mathbf{w}_{\phi_k}^T\mathbf{u}_{\phi_l}\mathbf{u}_{\phi_l}^T\mathbf{w}_{\phi_k}\stx(nT_s-\tau_{l}-pT_p)+
\zeta[\cdot], 
\end{align}
\normalsize
Subsequently, we match filter/correlate the receiver's fast-time radar data with the transmit signal over every consecutive pair of packets to obtain
\par\noindent\small
\begin{align}
\label{eq:mf1}
\mathbf{Y}^k[p] = \frac{1}{N}\left( \srx[n,p]\ast \Ga[-n]\right), p = 0, 1 \cdots, P, 
\end{align}
\normalsize
where $\mathbf{Y}^k[p] = [\mathbf{y}^k_1 \cdots \mathbf{y}^k_{N}]$ corresponds to the collection of $N$ range spectrums as displayed in Fig.~\ref{fig: block_diagram} for each $k^{th}$ beam. We obtain the range spectrums across all the packets and then detect the targets, corresponding to range index $\hat{n}$ with amplitude $\hat{a}$, using the detection strategy that will be discussed shortly. 
{\color{black} Subsequently, we implement MUSIC across $P$ packets for every $\hat{n}_b^{th}$ row of $\mathbf{Y}^k$ to estimate the Doppler velocity of each target. The algorithm is implemented as follows:
   
   First, we obtain the  autocorrelation matrix $\mathbf{R^k_y}[\hat{n}_b]$ as $\mathbf{y}^k[\hat{n}_b](\mathbf{y}^k)^H[\hat{n}_b]$ across every $\hat{n}_b^{th}$ cell of $\mathbf{Y}^k$. Since, $\mathbf{R^k_y}[\hat{n}_b]\in \mathbb{C}^{P \times P}$ is also a Hermitian matrix, all of its $P$ eigenvectors $[\mathbf{q_1, \cdots q_P}]$ are orthogonal to each other. Thus, if the eigenvalues of $\mathbf{R_y}$ are sorted in descending order, the eigenvector $\mathbf{q}_1$ corresponds to the largest eigenvalue that spans the signal subspace {\color{black} {\color{black} $\mathbf{{a}}_{sig}$}. The remaining $P-1$ eigenvectors correspond to the noise subspace, {\color{black} $\mathbf{{A}}_{noise}$}, and are orthogonal to {\color{black} $\mathbf{{a}}_{sig}$} (assuming that there is a single target for every range cell). When there are multiple $M$ targets for every range cell, then we can isolate $M$ eigenvalues for signal/target subspace, assuming that $M$ is always less than $P$ while the remaining $P-M$ eigenvalues correspond to noise subspace.
   
 Now, $\mathbf{e}[f_D]= [1\;e^{-j2\pi f_D T_p} \cdots e^{-j 2\pi (D-1)f_D T_p}]$ is the Doppler delay vector for the Doppler frequency $f_D \in \mathbb{R}^{1 \times D}$  and $T_p$ corresponds to the pulse repetition interval (PRI). We span $f_D$ from $-f_{D_{max}}$ to $-f_{D_{max}}$ with a resolution of $\Delta f_D$. Since this Doppler delay vector resides in the signal subspace {\color{black} $\mathbf{{a}}_{sig}$}, $\mathbf{e}[f_D]$ must be orthogonal to the noise subspace,  $\mathbf{e}[f_D] \perp {\color{black}\mathbf{{A}}_{noise}}$, which means $\mathbf{e}[f_D]$ is perpendicular to all the eigenvectors that span the noise subspace. In order to measure the degree of orthogonality of $\mathbf{e}[f_D]$ with respect to all the eigenvectors which belong to {$\mathbf{{A}}_{noise}$}, MUSIC algorithm defines a squared norm given as
 
 \par\noindent\small
\begin{align}
||{\color{black} \mathbf{{A}}_{noise}}^H\mathbf{e}[f_D] ||^2 = \mathbf{e}^H[f_D]{\color{black} \mathbf{{A}}_{noise}\mathbf{{A}}_{noise}^H}\mathbf{e}[f_D].
\end{align}
\normalsize
If $\mathbf{e}[f_D] \in  {\color{black}\mathbf{{a}}_{sig}} $, then  $||{\color{black}\mathbf{{A}}_{noise}^H}\mathbf{e}[f_D] ||^2 = 0$ as implied by the orthogonality condition.  The reciprocal of the squared norm expression is the pseudo-spectrum with sharp peaks at the Doppler frequencies. We identify the Doppler frequency of each $b^{th}$ MU for the $k^{th}$ beam from the corresponding frequency estimation of the peak of the MUSIC pseudo-spectrum as shown below

\par\noindent\small
\begin{align}
\label{eq:PseudoSpectrum}
\hat{f}_{D_b}^k = \argmax_{f_D} \frac{1}{\mathbf{e}^H[f_D]{\color{black} \mathbf{{A}}_{noise}\mathbf{{A}}_{noise}^H}\mathbf{e}[f_D]}.
\end{align}
\normalsize }

The above RSP procedure is repeated for every $k^{th}$ beam. Then, based on the detection of the presence of one or more mobile targets, we identify a subset of $\tilde{K}$ beams out of the total $K$ possible beams that should be subsequently used for communications. 
Note the following: (1) In the above discussion, we exclude those beams that indicate only static targets since these are regarded as clutter. (2) Static communication users are also excluded from the discussion since the optimal beam search only gets initiated when an MU moves out of a beam. (3) Beams corresponding to mobile communication users are identified provided they are detected by the radar. Based on the signal-to-noise ratio (SNR) levels, some of the MUs may remain undetected and, hence, go unserviced. (4) Also, some false targets may be detected. However, the beams corresponding to these false targets will be subsequently dropped using the UCB algorithm that follows. (5) More than one beam may be identified for the same MU since we consider both direct and multipath-based propagation paths. Therefore, when we have a very large number of beams indicating the presence of targets, we eliminate those beams where the range is close to the maximum unambiguous range, and the amplitude is just above the threshold to eliminate longer multipath. (6) Finally, based on the Doppler estimate of a mobile target in a specific $k^{th}$ beam along $\phi_k$, we estimate the velocity of the target. This estimate is used to compute the duration after which it is likely that the target will leave the beam assuming that it is moving at uniform velocity. Alternatively, we may be able to infer that a target has left the beam when the SNR becomes low. So, we compute $T_{\infty}$ based on the lower value of these two estimates. 
\subsection{Communication received signal and processing}  
The communication downlink signal along beam $\phi_k$ is an entire data packet $\mathbf{x}_{tx}$ of which $\stx$ is the preamble. The data packet is encoded, QAM modulated, upconverted to the carrier frequency, and transmitted through the ULA at the BS-TX. The received signal at the MU, 
\par\noindent\small
\begin{align}
\mathbf{x}_{rx}^k[n,p] = \mathbf{w}_{\theta}^T \mathbf{u}_{\theta} \mathbf{u}_{\phi_k}^T \mathbf{w}_{\phi_k} \mathbf{x}^k_{tx}\left(nT_s - \frac{\tau_b}{2}\right) e^{j\pi f_b pTp} + \\\nonumber\sum_{m=1}^M \mathbf{w}_{\theta_k}^T\mathbf{G}_m\mathbf{w}_{\phi_k} \mathbf{x}^k_{tx}(nT_s-\chi_{m})e^{j\pi f_{m}pT_p} +\\ \zeta'[\cdot],
\end{align}
\normalsize
is digitized and downconverted, demodulated, and decoded. 
Here $\mathbf{w}_{\theta} \in \mathbb{C}^{Q' \times 1}$ and $\mathbf{u}_{\theta} \in \mathbb{C}^{Q' \times 1}$ are the antenna weight vector and steering vector at the $Q'$ element array at the MU, and $\zeta'$ is the complex additive white Gaussian noise. $\mathbf{G}_m$ models the multipath propagation from BS to MU and $\chi_{m}$ corresponds to the range induced delay due to multipath. The downlink signal is processed at the MU and feedback about the communication link metrics are transmitted from the MU to the BS. \textcolor{black}{Note that the communication signal is distinguished from the radar signal at the BS receiver through cross-correlation with the Golay sequence. Due to the nature of the Golay sequence, the peak-to-sidelobe ratio after cross-correlation for the correct radar signal is very high compared to the communication signal. This has been discussed in greater length in the following subsection. }
\subsection{Radar detection strategy}
For each $k^{th}$ beam, the BS-RX may potentially receive one or more target scattered radar signals, communication uplink signals as well as noise. It is important that the BS-RX distinguishes between these components in order to detect the presence of a mobile target. In our case, we detect the presence of a target by measuring the peak-to-sidelobe level of the range spectrum obtained after the match filtering step shown in \eqref{eq:mf1}. In the absence of a target and clutter, the received signal will consist of
\par\noindent\small
\begin{align}
\mathbf{Y}^k[p] = \frac{1}{N}\left(\zeta[n,p]\ast \Ga[-n]\right).
\end{align}
\normalsize
The resulting range spectrum will show almost negligible peak to sidelobe level due to the mismatch between the Golay sequences and noise as shown in Fig.~\ref{fig:correlation_output}. Similarly, if the received signal is a communication uplink signal, then the Golay sequence in the preamble of the uplink signal will be distinct from the sequence used in the radar transmit signal. The mismatched signals poorly correlate resulting again in poor peak to sidelobe levels, which indicate the absence of a target as shown in Fig.~\ref{fig:correlation_output}. On the other hand, the peak to sidelobe level between $\stx$ and $\srx$ is $\sim 27$ dB.
\begin{figure}[htbp]
\vspace{-5mm}
\centering
\includegraphics[scale = 0.4]{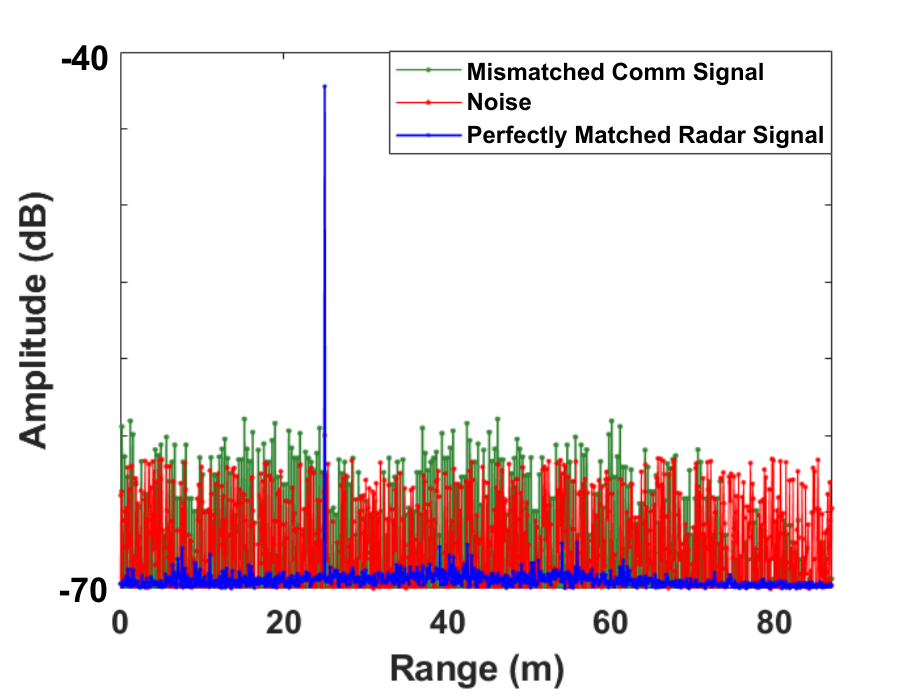}
\caption{\footnotesize Correlation output of $\stx$ with a perfectly matched radar signal, noise, and the mismatched preamble of the uplink communication signal.} 
\label{fig:correlation_output}
\end{figure}

\section{ISAC MAB Algorithm}
\label{sec:algos}
In this section, we set up the downlink beam-selection problem between BS and MU as an MAB. The BS aims to identify and communicate over the optimal beam, $\overline{k}$, with the highest SNR by ideally requiring the exploration of sub-optimal beams to be as low as possible. Once we identify $\overline{k}$ via \textit{sufficient exploration} of all beams, then the BS will \emph{exploit} this information by servicing the MU over $\overline{k}$ as often as possible. In this scenario, the challenge is that $\overline{k}$ is unknown and may change due to dynamic environmental conditions or due to the motion of the MU. Furthermore, with mmW communications, there are a large number of candidate beams ($K$), resulting in high exploration time. The first main objective of the proposed work is to use radar information to reduce the number of beams from $K$ to $\tilde{K}$ that must be \textit{explored} (stationary case). The second objective is to estimate the duration after which the beam search procedure must recommence due to dynamic environmental conditions (quasi-stationary case). In our discussion henceforth, we denote the set of the indices of the $K$ and $\tilde{K}$ beams to be $\mathbb{\mathbf{K}}$ and $\mathbb{\mathbf{\tilde{K}}}$.
\subsection{Stationary Case}
Figure \ref{fig:timing} depicts time-slotted communication, indexed by $t=0, T, 2T \cdots $ up to \emph{time horizon} ($T_{\infty}$). The BS selects a single $k^{th}$ beam in each $t^{th}$ time slot for transmission of a data packet and receives the reward - the SNR of the communication link, $\snr$, over the selected beam measured by the MU and send to the BS over the uplink. Due to the nature of the propagation channel, the 
$\snr$ across multiple beams are stochastically modeled using Gaussian distributions that are stationary and independent across beams. We define regret, $\Upsilon$ as the performance metric which is equal to the difference between the SNR of the optimal beam, $\snri$ and $\snr$ as shown in \par\noindent\small
\begin{equation}
\label{Eq:regret}
    \Upsilon =   T_{\infty} \snri  - \mathop{\mathbb{E}} \Bigg[ \sum_{k\in \K} \snr N_k \Bigg],
\end{equation} \normalsize
where $N_k$ is the number of times BS selects the beam $k$. The expectation is with respect to the random number of pulls of the beam $N_k$. Thus, the regret can be minimized by selecting $\overline{k}$, i.e., the beam with the highest SNR as many times as possible in $T_{\infty}$ through the UCB algorithm discussed below.
\subsubsection{SNR Based Beam Selection using UCB}
In conventional MAB algorithms, \textcolor{black}{all $K$ beams are selected once initially in the round-Robin manner.} Thereafter, under the principle of \textit{optimism under uncertainty}, the algorithm strategically selects the optimal beam as often as possible along the horizon. In \ref{alg:UCBSINR}, we have given the pseudo-code of the $\mathbf{UCB}_{SNR}$. As shown in Fig.~\ref{fig:timing}a, BS transmits the data frame over the selected beam in the downlink direction toward MU in each time slot.  In the first $K$ time slots \textcolor{black}{(round-Robin phase)}, BS selects each beam once (lines 4-6), while in the rest of the time slots ($t > (K-1)T$) \textcolor{black}{(regret-minimization phase)}, BS selects the beam with the highest UCB quality factor (lines 7-8), which is calculated based on the SNR feedback received in the previous time slots. During each time slot, the BS awaits the feedback via a UL acknowledgment to arrive from the MU, from which the SNR is estimated. We refer to this SNR as a reward for the selected beam. {\color{black} Although, as discussed in various works, MAB algorithms can be adapted to use other rewards or combinations, including power consumption, coding rate, antenna pattern, etc.,  \cite{modi2019transfer,MABoreward1,MABoreward2,MABoreward3}.  SNR has also been used as a reward in various published works such as \cite{SNRReward4,SNRReward3,MABoreward1,SNRReward1,modi2019transfer,SNRReward2}.} The slot duration, $T$, must thus be long enough to include the DL and UL packets, their propagation times, and the DL processing time at the MU. Naturally, the absence of an acknowledgment during the time slot implies the absence of a mobile target or a very poor wireless propagation environment. We denote the index of the selected beam and corresponding reward, i.e., instantaneous normalized SNR, as $I_t$ and $W_t$, respectively. At the end of each time slot, the parameters of all beams are updated (line 12). The UCB quality factor, \par\noindent\small
\begin{align}
\label{equ:UCB}
    UCB_k(t) = \frac{
    \hat{S}_k}{N_k} +\sqrt{\frac{2\log(t)}{N_k}},
\end{align}\normalsize
is calculated for each beam separately. 
Here, \textcolor{black}{ $\hat{S}_k$ denotes the cumulative reward, i.e., SNR, obtained over the $k^{th}$ beam till time slot $t$ and $N_k$ denotes the number of times the beam $k$ has been selected till time slot $t$. Thus, the ratio, $\frac{\hat{S}_k}{N_k}$, denotes the estimate of average SNR over beam $k$.}
The first term increases the UCB factor for the beams with high rewards in the past, while the second term ensures that all beams have been explored a sufficient number of times. The expected regret of $\mathbf{UCB}_{SNR}$ scales as $\mathcal{O}(\sum_{k \in \K\backslash \overline{k} }\frac{\log T}{\Delta_k})$ \cite{bubeck2012regret} where $\Delta_k=\snri-\snr$ for all $k\neq \overline{k}$. For fixed $T_{\infty}$, the distribution-independent upper bound is of the order $\mathcal{O}(\sqrt{KT_{\infty}})$ \cite{bubeck2012regret}. The $\mathbf{UCB}_{SNR}$ suffers from high exploration time and significant degradation in regret, especially when $\K$ is large. Furthermore, computational complexity and time taken to select the beam in each time slot scales linearly with $K$. Both these drawbacks limit the usefulness of $\mathbf{UCB}_{SNR}$ for mmW communication with a large number of narrow directional beams.
\begin{algorithm}[!ht] 
	\renewcommand{\thealgorithm}{Algorithm 1}
	\floatname{algorithm}{}
	\caption{$\mathbf{UCB}_{SNR}$: SNR Based Beam Selection}
	\label{alg:UCBSINR}
	\begin{algorithmic}[1]
		\STATE \textbf{Input:} $\K, T$

  \STATE \textbf{Initialize:} $N_k\leftarrow 0$ and $SNR^k\leftarrow 0$ for all $k$
		\FOR{$t=1, 2 \ldots T, $}
		    \IF {$t \leq \K$}  
        
		    \STATE Select beam, $I_t = t$.
     \hspace{10mm}{{\color{black}$\vartriangleright$(Round-Robin phase)}}\\
		    \ELSE 
		    \STATE $\forall k \in [\K]:$ compute $UCB_{k}(t)$ as given in Eq.~(\ref{equ:UCB})
		    
			\STATE Select beam, $I_t = \argmax\limits_{k \in [K]} UCB_{k}(t)$\\
        \hspace{30mm}{{\color{black}$\vartriangleright$(Regret-minimization phase)}}\\
			\ENDIF
			\STATE BS transmits a data frame over beam $I_t$ 
   \STATE MU observes instantaneous normalized SNR, $W_{t}$
		and communicate to BS over the uplink.
			\STATE $N_{I_t} \leftarrow N_{I_t}+1$ and $S_{I_t}\leftarrow S_{I_t} +W_{t}$.
		\ENDFOR 
	\end{algorithmic}
\end{algorithm}
\begin{figure}[!t]
\includegraphics[scale = 0.45]{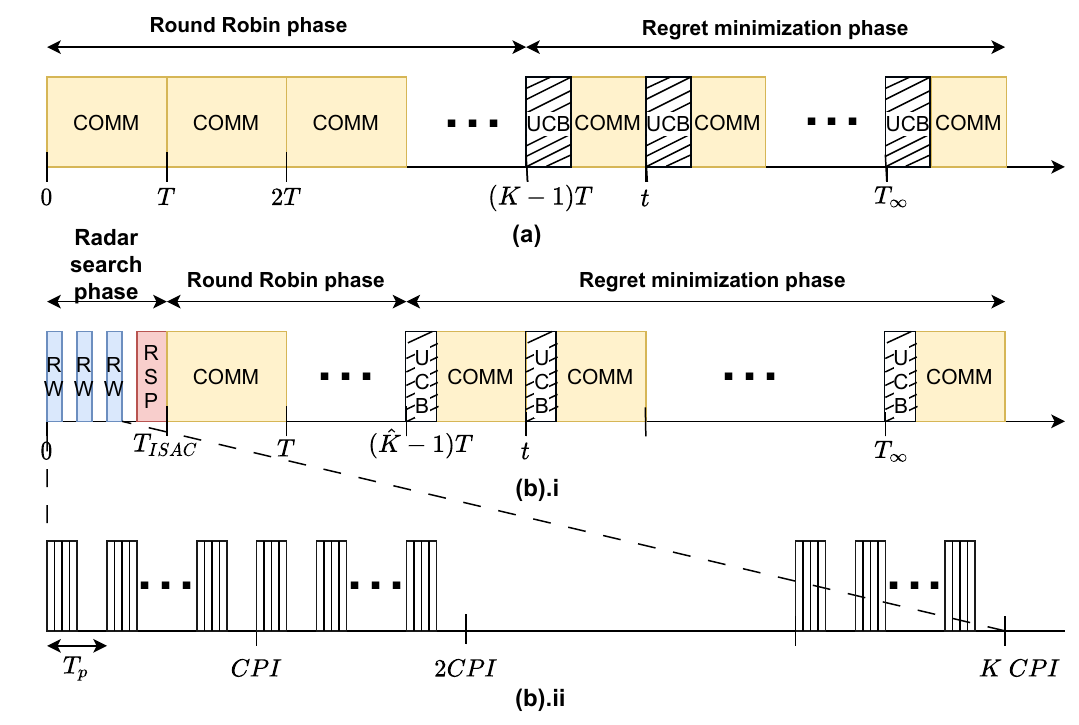}
\caption{\footnotesize \textcolor{black}{(a) Timing diagram for beam selection for communications (COMM) based on UCB algorithm (b).i. Timing diagram for beam selection for proposed ISAC framework using radar waveform (RW), radar signal processing (RSP) and UCB, (b).ii. Radar waveforms with 50\% duty cycle for multiple pulse repetition interval $T_p$ over coherent processing interval (CPI). }}
\label{fig:timing}
\end{figure}
\subsubsection{Proposed MAB for ISAC}
 In the proposed  $\mathbf{UCB}_{ISAC}$ approach, we reduce the number of beams from $K$ to $\tilde{K}$ using the RSP in ISAC described in the previous section, {\color{black} thereby reducing the distribution-independent upper bound to order $\mathcal{O}(\sqrt{\tilde{K}T_{\infty}}$).} As shown in Fig.~\ref{fig:timing}(b), we used the RSP, in the beginning, to identify the beams with targets/scatterers, their strength, and the type (stationary or moving). Since the radar pulses (extracted from the communication packet preamble) are significantly shorter than the communication data frame, and RSP does not need frame decoding at MU and UL communication, the expected duration of the {\color{black}radar search phase}, $T_{ISAC}$ is significantly smaller and can be completed within a few time slots depending on the desired accuracy. In the {\color{black}radar search phase}, each time slot can be divided into $K$ sub-slots each of CPI duration as shown in Fig.~\ref{fig:timing}b.ii. Within each CPI, multiple ultrashort radar pulses with $T_p$ pulse repetition intervals are transmitted to detect the presence of one or more targets and estimate their Doppler velocity as described in the previous section. The set of beams where mobile targets are detected constitutes $\tilK$. After the {\color{black}radar search phase}, we use UCB with $\tilde{K}$ beams. Our main hypothesis is that $\tilde{K}$ is likely to be much lower than $K$ in a low-clutter environment, resulting in an overall shorter \textcolor{black}{round-Robin phase and regret-minimization phase}. The smaller number of beams results in lower regret and lower computation time. Both allow a higher number of selections of an optimal beam and more time for data communication in a given slot as shown in Fig.~\ref{fig:timing}b.i when compared to Fig.~\ref{fig:timing}a. {\color{black}Later in Section \ref{sec:hardware}, we compare the execution times of the UCB and RSP algorithms on SoC and demonstrate that the RSP augmented with UCB offers superior performance over conventional UCB. This analysis is novel when compared to existing literature and critical to validate the superiority of the proposed approach over conventional MAB approaches.} The pseudo-code of the  $\mathbf{UCB}_{ISAC}$ is shown in \ref{alg:UCBISAC}. First, RSP is performed (Lines 3-5) using the algorithms discussed in the previous section and the indices of the beams selected are stored in $\beta_{\tilK}$. Subsequently, BS selects the beam $Q_t$ with the highest value of UCB (Line 9-10). Since the RSP provides a subset of the total beams ($\tilK \in \K$) where a mobile target may be present, the overall time of \textcolor{black}{round-Robin phase and regret-minimization phase} is reduced.
\begin{algorithm}[!ht] 
	\renewcommand{\thealgorithm}{Algorithm 2}
	\floatname{algorithm}{}
	\caption{$\mathbf{UCB}_{ISAC}$: ISAC RSP and UCB Based Beam Selection}
	\label{alg:UCBISAC}
	\begin{algorithmic}[1]
		\STATE \textbf{Input:} $\K, T, T_{ISAC}$
		\FOR{$t=1, 2 \ldots T, $}
                \IF {$t \leq T_{ISAC}$}
		    \STATE Find ${\tilK}$ potential beams out of $\K$  using ISAC and store their indices in $\beta_{\tilK}$  
	         \STATE \textbf{Initialize:} $N_{\tilK}\leftarrow 0$ and $S_{\tilK}\leftarrow 0$ for all $\tilK$ \\\hspace{30mm}{$\vartriangleright$(Radar search phase)}\\
              \ELSIF {$t \leq T_{ISAC}+\tilK$}
		    \STATE Find $Q_t =t-T_{ISAC}$ \\
           \hspace{30mm}{$\vartriangleright$(Round robin phase)}\\
		    \ELSE 
		    \STATE $\forall k \in [{\tilK}]:$ compute $UCB_{k}(t)$ as given in Eq.~(\ref{equ:UCB})
			\STATE Find $Q_t =\argmax\limits_{k \in [\tilK]} UCB_{k}(t)$ \\
           \hspace{30mm}{$\vartriangleright$( Regret-minimization phase)}\\
			\ENDIF
   \STATE Select beam with index, $I_t=\beta_{Q_t}$ 
			\STATE BS transmits a data frame over beam $I_t$ 
   \STATE MU observes instantaneous normalized SNR, $W_{t}$
		and communicate to BS over the uplink.
			\STATE $N_{Q_t} \leftarrow N_{Q_t}+1$ and $S_{Q_t}\leftarrow S_{Q_t} +W_{t}$.
		\ENDFOR 
	\end{algorithmic}
\end{algorithm} 

{\color{black}Compared to $\mathbf{UCB}_{SNR}$, $\mathbf{UCB}_{ISAC}$ potentially offers lower regret due to the following reasons: 1) Faster target detection: The identification of the presence of targets using radar is significantly faster since returns of the scattered signals from the short-range targets are nearly instantaneous with a short round-trip delay of the order of a few $ns$. On the other hand, for a communication signal such as 5G, one slot is at least 4 $ms$ assuming downlink sub-frame (1 $ms$), uplink sub-frame for reward feedback (1 $ms$), downlink (1 $ms$) and uplink data processing (1 $ms$).
   2) The proposed algorithm focuses on a subset of the total beams, $\tilde{K}$ which is likely to be much lower than $K$ in a low-clutter environment, resulting in an overall shorter duration of \textcolor{black}{round-Robin phase and regret-minimization phase}. The smaller number of beams results in lower regret and lower computation time, allowing a higher number of selections of an optimal beam and more time for data communication in a given slot as shown in Fig.~\ref{fig:timing}b.i when compared to Fig.~\ref{fig:timing}a.
To quantify this gain, let us consider a bandit instance. The radar identifies a set of beams, $\tilK$, which is a random set depending on the distribution of scatterers. We can assume $\tilK$ includes the optimal beam, $\overline{k}$, as this beam is likely to produce a strong scattered signal at the BS-RX due to the high SNR and radar is unlikely to miss the MU. Hence, the optimal arm is the same in any realized set $\tilde{{K}}$. Expected regret over the set $\tilde{{K}}$ is  $\mathcal{O}(\sum_{k \in \tilde{{K}}\backslash \overline{k}} \frac{\log T}{\Delta_k})$. Clearly this bound is smaller than  $\mathcal{O}(\sum_{k \in {{K}}\backslash \overline{k}} \frac{\log T}{\Delta_k})$ obtained for the previous case. Taking expectation over the random realizations $\tilde{{K}}$, we get expect regret of $\textbf{UCB}_{ISAC}$ as $\mathbb{E}\left[\mathcal{O}(\sum_{k \in {\tilde{{K}}}\backslash \overline{k}} \frac{\log T}{\Delta_k})\right] \leq \mathcal{O}(\sum_{k \in {{K}}\backslash \overline{k}} \frac{\log T}{\Delta_k})$. Thus, $\textbf{UCB}_{ISAC}$ is better than that of $\mathbf{UCB}_{SNR}$, resulting in an improvement in the performance.}  
\subsection{Quasi-Stationary Case}
Next, we consider quasi-stationary environmental conditions where channel and/or target parameters may change over time in a manner that is neither periodic nor known. The conventional approach to adapt the MAB algorithm in a quasi-stationary environment is to restart the algorithm after periodic time intervals. Here, the restart interval should be large enough to \textit{sufficiently explore} all beams and a large number of beams puts a lower bound on the restart interval. Thus, this approach does not work well when the minimum exploration time is higher than the restart time. Another approach is based on the \textit{change detection method} where the beam/statistics over the different time windows are compared, and the algorithm is restarted whenever the difference is above a certain threshold for any or few sets of beams \cite{CDMAB1,CDMAB2}. This approach demands huge memory and computation costs to store the learned estimation of all beams at different window intervals. Furthermore, the threshold selection is not trivial since it depends on deployment conditions. In this work, we proposed a novel \textit{change detection method} using RSP of ISAC, which addresses both of the above issues without additional hardware costs. 

The pseudo-code of the \textit{change detection method} is shown in \ref{alg:quasi_ISAC}. The RSP provides us with two-dimensional position ($r_b$) and Doppler velocity ($v_b=\frac{\lambda f_b}{2}$) estimates of a $b^{th}$ MU for each $\phi_k$. Further, the optimal beam $\phi_{\overline{k}}$ is obtained through UCB learning.  However, the heading angle or direction of the velocity is not known. Hence, we adopt a pragmatic approach to estimate the duration of the target within the beam by assuming that the MU is headed in a lateral direction (along $x$ axis) as shown in Fig.~\ref{fig:target model}(b) with uniform lateral velocity ($v_b\cos\phi_{\overline{k}}$). The actual durations of the MU within a beam are likely to be much longer if it is headed closer to a radial direction and shorter if the MU is headed closer to a tangential direction with respect to the BS. In the former case, the $\snri$ of the previous time slot corresponding to the optimal beam is likely to remain steady for a long time, while in the latter case, the $\snri$ of the previous time slot is likely to deteriorate quickly. Hence, the decision to restart the beam realignment process is carried out in conjunction with an estimate of $T_{\infty}$ along with the $\snri$. For instance, if MU follows a lateral motion, $T_{\infty}$ corresponds to the time when MU leaves the optimal beam $\phi_{\overline{k}}$ of beamwidth $\Delta \phi$. Thus, $T_{\infty}$ can be calculated as $T_{\infty} = \frac{r_b\Delta \phi}{v_b\cos\phi_{\overline{k}}}$.
\begin{algorithm}[!ht] 
	\renewcommand{\thealgorithm}{Algorithm 3}
	\floatname{algorithm}{}
	\caption{ Change detection method : $T_{\infty}$ Using RSP for ISAC}
	\label{alg:quasi_ISAC}
	\begin{algorithmic}[1]
		\STATE \textbf{Input:} $\phi_{k}, T$
		\STATE \textbf{Output:} $T_{\infty}$ 
		    \STATE Find $f_b, r_{b}$  through RSP for each $\phi_k$ 
              \STATE Find 2D position coordinates of MU using $r_b$ and $\phi_{\overline{k}}$
              \STATE Find $T_{\infty}$ based on $v_b\cos\phi_{\overline{k}}, r_b$ and $\Delta \phi$ using geometry.
	\end{algorithmic}
\end{algorithm}
{\color{black} For regret analysis, we compare the regret of $\mathbf{UCB}_{UCB}$ and $\mathbf{UCB}_{ISAC}$ assuming there are $\mathcal{C}$ change points at the fixed interval of $T_{\infty}$. We assume that the distributions of all beams are changed at every change point. We do not need \ref{alg:quasi_ISAC}. and simply restart both algorithms after every $T_{\infty}$ slot. The expected regret of $\mathbf{UCB}_{SNR}$ and $\mathbf{UCB}_{ISAC}$ scale as $\mathcal{O}(\sqrt{\mathcal{C}KT_{\infty}})$ and $\mathcal{O}(\sqrt{\mathcal{C}\tilde{K}T_{\infty}})$, respectively. Next, we assume that the $\mathcal{C}$ change points are uniformly distributed over the horizon and are unknown. In this case, $\mathbf{UCB}_{SNR}$ needs an additional change detection algorithm \cite{CDMAB2} while $\mathbf{UCB}_{ISAC}$ needs simple computations discussed in \ref{alg:quasi_ISAC}. Interestingly, the expected regret of $\mathbf{UCB}_{SNR}$ with change detection is of the order $\mathcal{O}(\sqrt{\mathcal{C}KT_{\infty}})$ as proved in \cite{CDMAB2}. However, this requires prior knowledge of a number of change points, $\mathcal{C}$. Using the same approach, we can show that the expected regret of $\mathbf{UCB}_{ISAC}$ is of the order $\mathcal{O}(\sqrt{\mathcal{C}\tilde{K}T_{\infty}})$ and better than $\mathbf{UCB}_{SNR}$. Additionally, the proposed approach does not need prior knowledge of $\mathcal{C}$. Empirically, the change detection in $\mathbf{UCB}_{ISAC}$ is much faster than  \cite{CDMAB2} as the $\mathbf{UCB}_{ISAC}$ uses the radar localization information for beam change prediction while \cite{CDMAB2} is based on detecting the beam change by observing the statistics of all beams via forced exploration at regular intervals of time. We validate this hypothesis using the simulation results presented in Section~\ref{sec:Results}. A corresponding closed-form expression for regret incorporating change detection time is a challenging and interesting direction for future work.  } 

\section{Simulation Setup}
\label{sec:Simulationsetup} 
In this work, we consider a three-dimensional Cartesian space where $xy$ forms the ground plane and $z$ is the height axis. We locate the BS-TX/RX at $[0,0,0]$ m with a uniform linear array (ULA) of 32 antennas with half-wavelength antenna spacing and center frequency $f_c$ of 60 GHz. The total number of pre-determined beams is 41, spanning from -80$^o$ to -80$^o$. The MU-TX/RX also consists of a 32-element ULA with half-wavelength spacing. The Doppler resolution obtained from the MUSIC algorithm is 400 Hz, which corresponds to a velocity resolution of 1 m/s. The ISAC radar is characterized by 100\% probability of detection and 0.48\% probability of false alarm.

We consider three different types of target/MU in both stationary and quasi-stationary scenarios.
Firstly, we model MU as a simple isotropic point target as shown in Fig.~\ref{fig:target model}(a) and (b). The second and third target is an extended target with multiple point scatterers corresponding to different parts of the targets' bodies. The second target is a pedestrian modeled with 27-point scatterers corresponding to ellipsoidal body parts as shown in Fig.~\ref{fig:target model}(c). The animation data for the walking pedestrian are obtained from motion capture technology and the average RCS is 0 dBsm \cite{ram2008simulation,ram2010simulation}. The third target is a mid-size car obtained from a freely available computer-aided design model, rendered with 6905 metallic plates, as shown in Fig.~\ref{fig:target model}(d). The car is animated using the techniques explained in \cite{pandey2020database,pandey2022classification}. The average RCS of the car is 10 dBsm. 

We consider two types of target trajectories. In the first trajectory shown in Fig.~\ref{fig:target model}(a), the target moves radially with respect to the BS with 3 m/s velocity over a duration of 8 seconds. Therefore, it does not move out of the beam during the entire duration. The channel characteristics also remain the same throughout the duration. Thus, this corresponds to a stationary scenario. In the second trajectory shown in Fig.~\ref{fig:target model}(b), the target moves laterally (along $x$ axis) from a starting position of $[20,15,0]$ m at 3 m/s over a duration of 8 seconds. Here, the MU moves out of the beam after a certain interval of time. In both these scenarios, we introduce $L$ static clutter scatterers (SCS) that are detected as radar targets as well as $M$ multipath components (MC) with non-zero Doppler frequency shifts. The radar signals are modeled with an SNR of 10 dB. 

We model OFDM-based multi-carrier PHY for the IEEE 802.11ad protocol based on the specifications provided in Table~\ref{tab:specs}. 
\begin{figure}[htbp]
    \centering
    \includegraphics[scale=0.52]{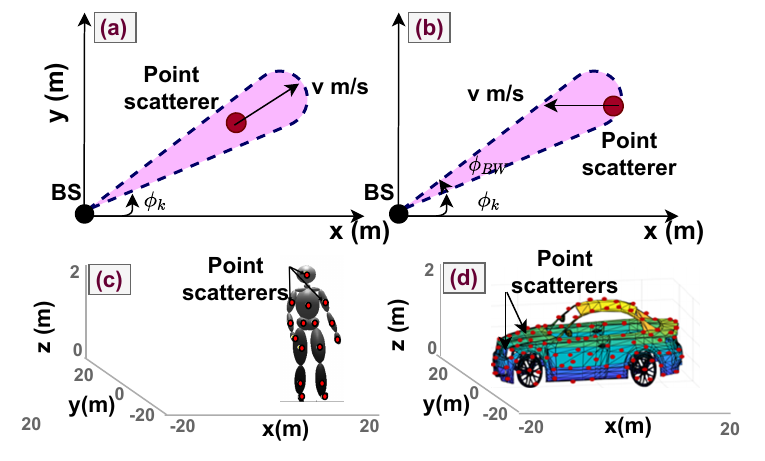}
    \caption{ \footnotesize (a) Radial trajectory followed by a point scatterer, (b) lateral motion along $x$ axis followed by a point scatterer.  multiple point scatterer model of (c) pedestrian and (d) mid-size car.}
    \label{fig:target model}
\end{figure}
\begin{table}[htbp]
\centering
\caption{\footnotesize Specifications of Wireless PHY}
    \begin{tabular}{p{5cm}|p{2cm}}
    \hline \hline
       \textbf{Specifications} & \textbf{Values} \\
    \hline \hline
  Carrier Frequency & 60GHz \\  
  Bandwidth & 1.76GHz \\
  Modulation Scheme & QAM \\ 
  Modulation Order & 16 \\ 
  Number of OFDM Subcarriers per Symbol   & 512\\ 
  Number of Data Subcarriers per Symbol   & 48\\ 
  Number of Data Bits per Subcarrier   & 42\\ 
    Number of OFDM Symbols  & 20\\ 
     
   Number of Null Subcarriers & 12 \\ 
   Number of Pilots & 4 \\ 
   Number of Antennas at BS/MU & 32 \\ 
    
    \hline 
\end{tabular}
\label{tab:specs}
\end{table}
The SNR for the downlink communications is varied from -10 dB to 10 dB in steps of 5 dB. For each SNR value, we perform Monte Carlo simulations with 15 independent experiments and present the average across all the SNRs as the output.

\section{Simulation Performance Analysis}
\label{sec:Results} 
We analyze the performance of the proposed $\mathbf{UCB}_{ISAC}$ algorithm and benchmark it with two conventional MAB algorithms - {$\mathbf{UCB}_{SNR}$ and lower upper confidence bound (LUCB) algorithm as described in \cite{kalyanakrishnan2012pac} - in terms of cumulative bit error rate (BER) and throughput. We also benchmark with digital beamforming (DBF) and the policy of random beam selection. DBF allows the fastest optimal beam selection since it enables the simultaneous evaluation of all the beams. Hence, the DBF results show us the best possible results (lowest BER) that can be realized in theory. However,  the implementation of multiple time-synchronized mmW channels is both prohibitively costly and challenging for most applications. Random beam selection without utilizing any type of learning, on the other hand, will give rise to the worst possible performance with the highest BER. We evaluate the performance of all the above-mentioned algorithms for two scenarios. First, is a stationary scenario where we consider MU moving in a radial direction with respect to BS as shown in Fig.~\ref{fig:target model}(a). The second scenario is quasi-stationary where MU is moving laterally along $x$ direction as shown in Fig~\ref{fig:target model}(b). 

\subsection{Stationary Scenario}
First, we consider stationary conditions where the MU remains within the beam over the entire time horizon and hence we do not have to estimate $T_{\infty}$. We analyze the effects of different target models, number of MCs, and radar parameters on the performance of the algorithms. \\

\noindent \textbf{Different target models:} First, we consider the simple point target model of the MU along with one MC and no SCS. We compare the cumulative BER of all the algorithms over 2000 time slots as shown in Fig.~\ref{fig:target analysis}(a). The total number of pre-determined beams is 41, spanning from -80$^o$ to -80$^o$. We observe that the $\mathbf{UCB}_{ISAC}$ algorithm performs better than both the conventional MAB algorithms and the random beam selection method because of the shorter \textcolor{black}{round-Robin phase and regret-minimization phase}. Next, we present the average throughput per time slot of each algorithm, which is calculated as $\left(1 - \hat{BER}\right)\frac{D}{T}$ where $\hat{BER}$ is the mean BER and $D$ is the total number of data bits in each packet which corresponds to 40,000 bits as per specifications mentioned in Table~\ref{tab:specs}. $T$ is fixed to 4 $ms$ for each slot.
In Table~\ref{tab:throughput_point}, we observe that the throughput of $\mathbf{UCB}_{ISAC}$ is higher than the conventional UCB algorithms. 
\begin{figure}[htbp]
    \centering
    \includegraphics[scale=0.25]{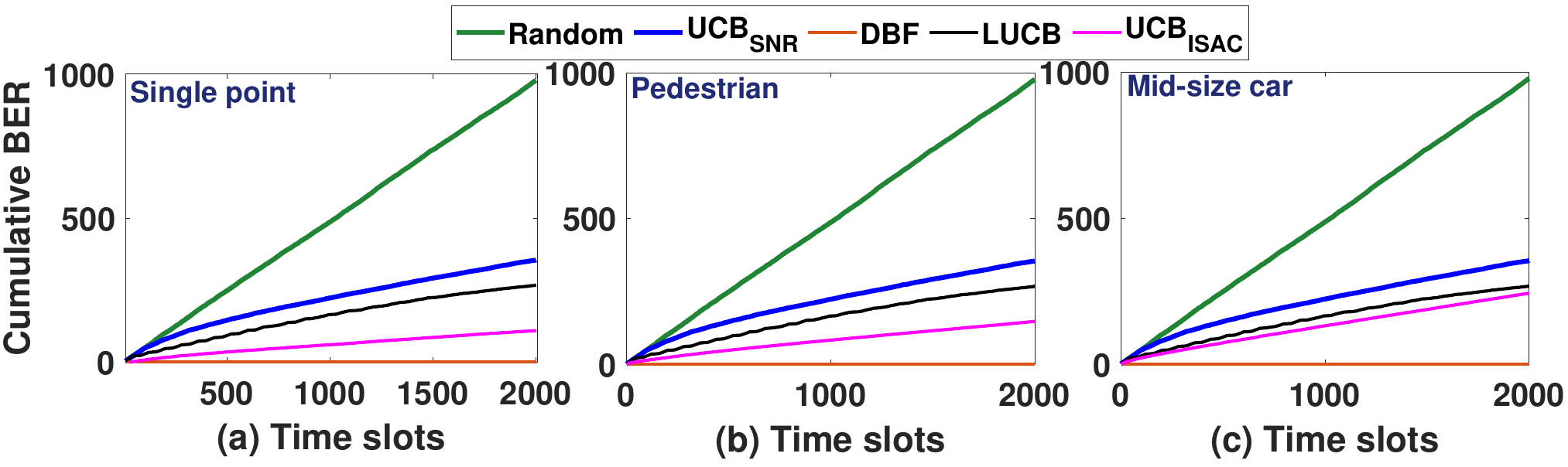}
    \caption{\footnotesize Cumulative BER at different time slots for 1 MU, 1 MC, and where MU is (a) single point target, (b) pedestrian and (c) mid-size car.}
    \label{fig:target analysis}
\end{figure}

\begin{table}[htbp]
\centering
\caption{ \footnotesize Comparison of Throughput of different algorithms and targets models in stationary case }
\begin{tabular}{|l|lll|}
\hline
{\textbf{Algorithms}} & \multicolumn{3}{c|}{\textbf{Throughput (Mbps)}}                     \\ \cline{2-4} 
                                     & \multicolumn{1}{l|}{\textbf{Single point}} & \multicolumn{1}{l|}{\textbf{Pedestrian}} & \textbf{Mid-size car} \\ \hline 
                                     $\mathbf{UCB}_{ISAC}$               & \multicolumn{1}{l|}{9.42}                  &  \multicolumn{1}{l|}{9.26}                    & 8.78                      \\ \hline
\textbf{Random}                      & \multicolumn{1}{l|}{5.10}                  & \multicolumn{1}{l|}{5.10}                    &                     5.10  \\ \hline
$\mathbf{UCB}_{SNR}$                & \multicolumn{1}{l|}{8.22}                  & \multicolumn{1}{l|}{8.22}   &  8.22                     \\ \hline
\textbf{LUCB} & \multicolumn{1}{l|}{8.63}   & \multicolumn{1}{l|}{8.63}                    &       8.63                \\ \hline
\textbf{DBF} & \multicolumn{1}{l|}{10}   & \multicolumn{1}{l|}{10}                    & 10                      \\ \hline
\end{tabular}
\label{tab:throughput_point}
\end{table}

Next, we consider the target as a pedestrian modeled with multiple-point scatterers along with one MC and no SCS. As shown in Fig.~\ref{fig:target analysis}(b), we observe that the cumulative BER of $\mathbf{UCB}_{ISAC}$ is slightly higher than that of a simple point target since the greater spatial extent of the pedestrian causes more beams to get selected. Consequently, throughput is reduced as shown in Table~\ref{tab:throughput_point} for $\mathbf{UCB}_{ISAC}$. Interestingly, there is no change in the performance of other algorithms where the difference between point and extended radar targets is not of significance to the algorithms. Lastly, we analyze the performance of the ISAC algorithm when we model the MU as a mid-size car that consists of 6905 point scatterers while keeping all the other environmental conditions identical. Here, we observe that the cumulative BER is further increased as observed in  Fig.~\ref{fig:target analysis}(c) as the car is much larger spatially resulting in a lower value of throughput as shown in Table~\ref{tab:throughput_point}. However, for all three targets, the proposed $\mathbf{UCB}_{ISAC}$ still performs better than $\mathbf{UCB}_{SNR}$ and LUCB.   \\

\noindent \textbf{Number of multipath components:} Next, we show the performance analysis of all the algorithms for different number of MC while keeping the other parameters the same as Fig.~\ref{fig:target analysis}(a). Cumulative BER  obtained for the different numbers of MC is presented in Fig.~\ref{fig:point analysis}(a). Here, we observe that the overall performance of all the algorithms degrades with the increase in the number of MC due to a greater number of candidate beams being selected with the increase in the number of MC. However, $\mathbf{UCB}_{ISAC}$ still performs better than the conventional MAB algorithms in each case.\\
\begin{figure}[htbp]
    \centering
    \vspace{-5mm}
    \includegraphics[scale=0.24]{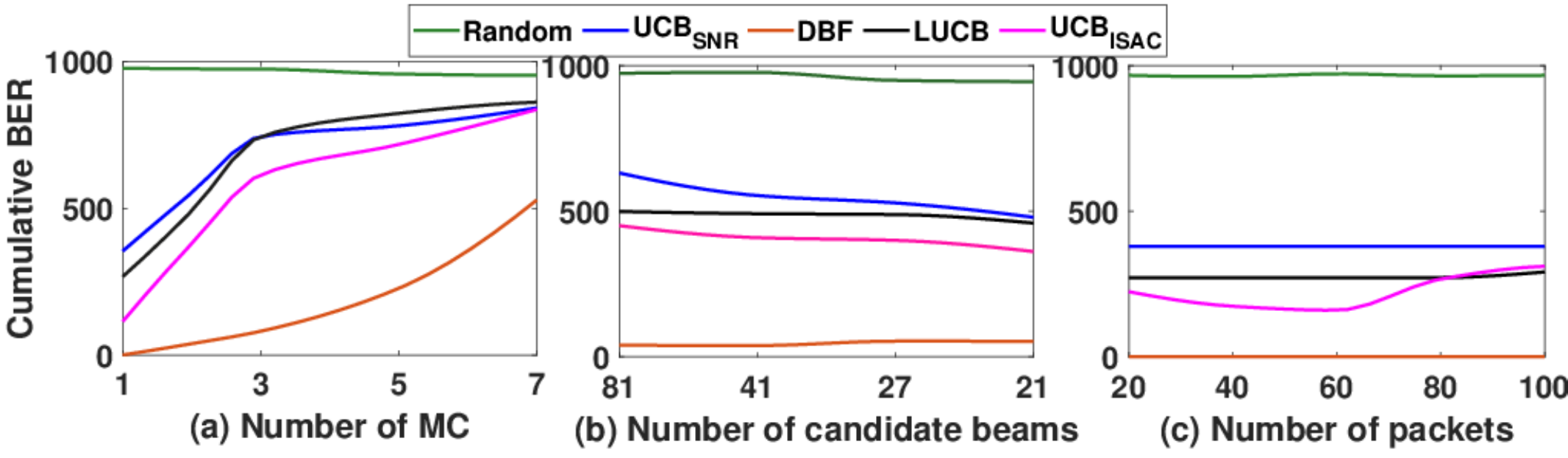}
    \caption{\footnotesize \textcolor{black}{Cumulative BER at the last time slot for (a) the different number of multipath components, (b) different number of candidate beams, and (c) the different numbers of radar packets.}}
    \label{fig:point analysis}
\end{figure}

{\color{black} \noindent \textbf{Different number of candidate beams} We also analyze the performance of all the algorithms for different numbers of candidate beams with the corresponding beamwidth through the antenna array at the base station. For instance, a beamwidth of $2^o$ results in a total of 81 candidate beams that span from $-80^o$ to $80^o$ while a beamwidth of $8^o$ results in 21 beams.  In the experiment, we assume MU as a simple point target and fix the number of multipath components (MC) to 2 and no static clutter scatterers (SCS). As shown in 
 Fig.~\ref{fig:point analysis}(b), we compare the cumulative BER at the end of the horizon for different numbers of candidate beams and observe that cumulative BER decreases as the number of beams reduces due to lower duration of \textcolor{black}{round-Robin phase and regret-minimization phase}, i.e.,  lower regret is offered for smaller beamwidth (i.e., large $K$). However, the proposed  $\mathbf{UCB}_{ISAC}$ algorithm significantly outperforms the other conventional approaches in all cases. 
}\\
\noindent \textbf{Integration factor:} Lastly, we observe the effect of radar data integration on the RSP with the rest of the parameters identical to Fig.~\ref{fig:target analysis}(a). We observe that initially as the number of radar packets increases, the regret of $\mathbf{UCB}_{ISAC}$ decreases in Fig.~\ref{fig:point analysis}(c) resulting in lower cumulative BER.  This is due to the improvement in the SNR with more radar packets. However, the latency also proportionately rises with increased RSP time. Thus, we observe an increase in the cumulative BER for a much larger number of radar packets.
\subsection{Quasi-stationary Scenario}
In this section,  we discuss the performance of all the algorithms in a quasi-stationary environment in which the beams initially selected by all the algorithms become non-optimum after a certain time when the MU exits the beam. Here, we estimate $T_{\infty}$ to restart the algorithms when the MU exits the beam. The fixed time slot interval chosen to restart $\mathbf{UCB}_{SNR}$ and LUCB algorithms is 2000. The cumulative BER is obtained over a longer time horizon comprising 8000 time slots.

Firstly, we discuss the case where MU is modeled as a simple point target with one MC and no SCS. In Fig.~\ref{fig:realign}(a), we observe that RSP is restarted at the exact time slot (425) when it goes out of the beam which results in a lower regret as MU is always within the optimum beam. On the other hand, due to the fixed time slot interval (2000) chosen for restarting $\mathbf{UCB}_{SNR}$ and LUCB algorithms, the BS persists longer with a non-optimum beam, resulting in a higher cumulative BER. We also compare with the case where the BS persists with initially selected beams throughout the time horizon and the beam search is not initiated at all. The results are shown in Fig.~\ref{fig:realign}(b). We observe a drastic increase in the cumulative BER of each algorithm once the beams selected initially are no longer optimum. 

Next, we present the results where we model MU as pedestrian and car for the same scenario shown in Fig.~\ref{fig:realign}(a) and present the cumulative BER in Fig.~\ref{fig:realign}(c) and (d). In both cases, the regret of $\mathbf{UCB}_{ISAC}$ increases as the number of beams has increased due to the extended target model. 

Further, we present the average throughput per time slot for each algorithm and target model in Table~\ref{tab:throughput_quasi}. We observe here that the throughput of $\mathbf{UCB}_{ISAC}$ is higher for all the target models as compared to the $\mathbf{UCB}_{SNR}$ and LUCB algorithms.   
\begin{figure}[htbp]
    \centering
    \includegraphics[scale=0.35]{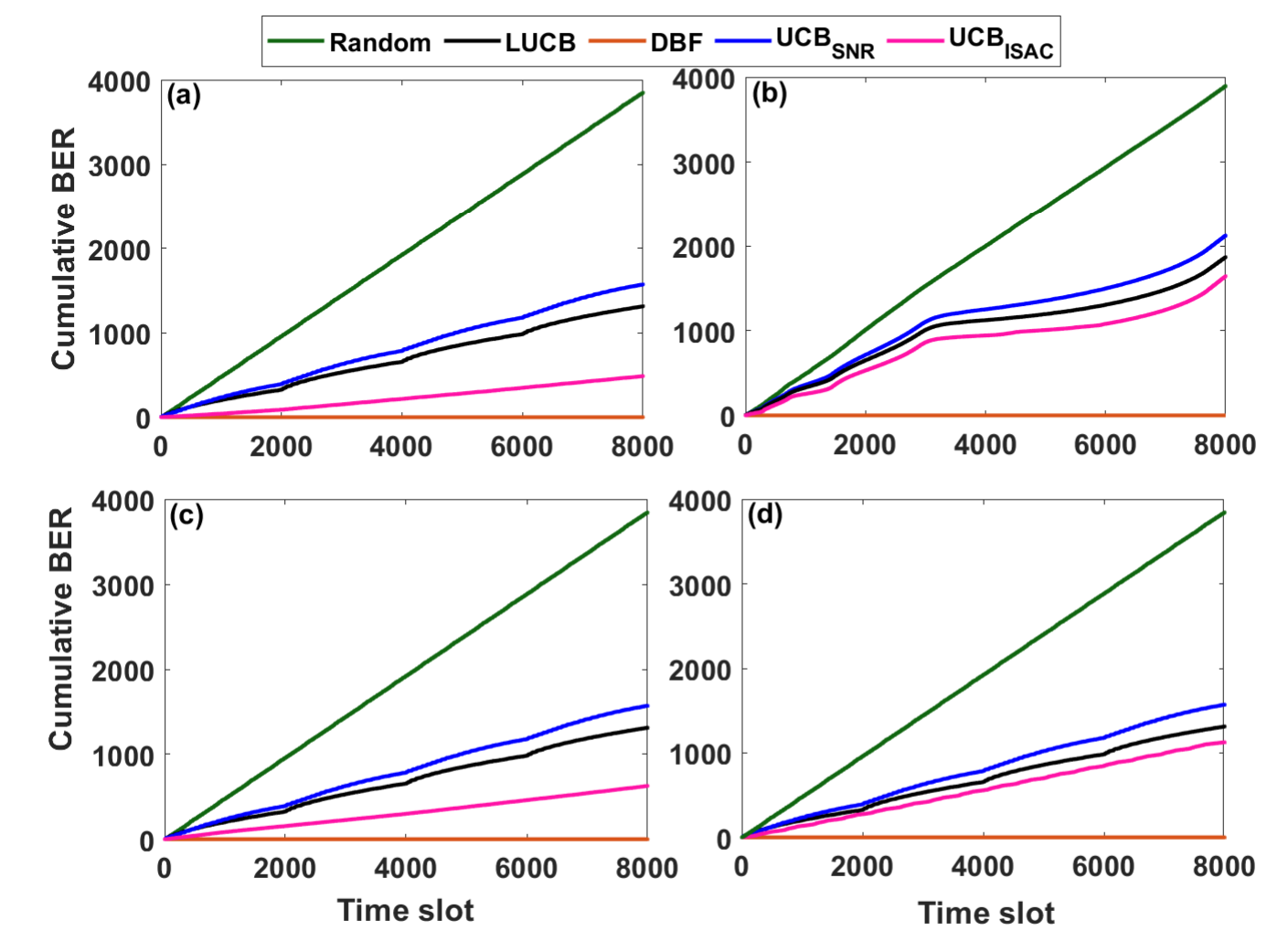}
    \caption{\footnotesize{\color{black}Cumulative BER at different time slots (a) with realignment and (b) without realignment for 1 point target MU and 1 MC. Cumulative BER at different time slots with realignment for (c) MU modeled as a pedestrian and (d) MU modeled as a mid-size car.}}
    \label{fig:realign}
\end{figure}

\begin{table}[htbp]
\centering
\caption{\footnotesize  Comparison of throughput of different algorithms and targets models in quasi-stationary case }
\begin{tabular}{|l|lll|}
\hline
{\textbf{Algorithms}} & \multicolumn{3}{c|}{\textbf{Throughput (Mbps)   }}                  \\ \cline{2-4} 
                                     & \multicolumn{1}{l|}{\textbf{Single point}} & \multicolumn{1}{l|}{\textbf{Pedestrian}} & \textbf{Mid-size car} \\ \hline 
$\mathbf{UCB}_{ISAC}$               & \multicolumn{1}{l|}{9.64}                  &  \multicolumn{1}{l|}{9.32}                    &  9.01                     \\ \hline
\textbf{Random}                      & \multicolumn{1}{l|}{5.24}                  & \multicolumn{1}{l|}{5.24}                    &                     5.24  \\ \hline
$\mathbf{UCB}_{SNR}$                & \multicolumn{1}{l|}{8.87}                  & \multicolumn{1}{l|}{8.87}   &  8.87                     \\ \hline
\textbf{LUCB} & \multicolumn{1}{l|}{8.89}   & \multicolumn{1}{l|}{8.89}                    &       8.89                \\ \hline
\textbf{DBF} & \multicolumn{1}{l|}{10}   & \multicolumn{1}{l|}{10}                    & 10                      \\ \hline
\end{tabular}
\label{tab:throughput_quasi}
\end{table}

Next, we examine the performance of all the algorithms for specific scenarios involving a larger number of candidate beams, and a large number of MC and the presence of SCS. In all the scenarios, the MU is modeled as a simple point target.\\
\begin{figure}[htbp]
    \centering
    \includegraphics[scale=0.1]{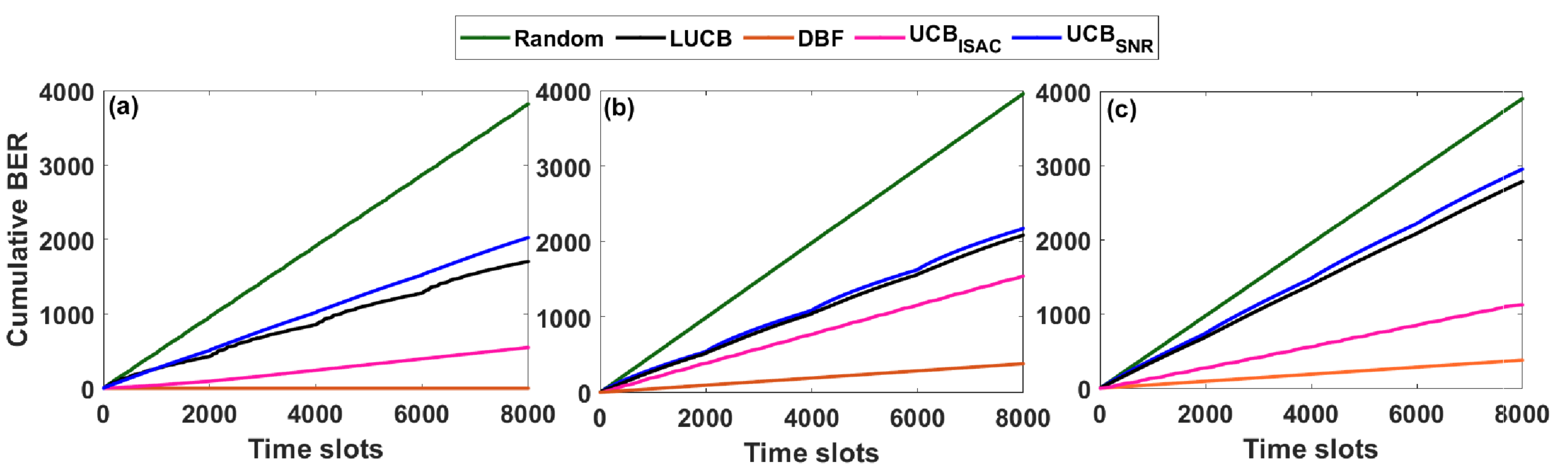}
    \caption{\footnotesize {\color{black} Cumulative BER at different time slots for 1 MU with (a) 81 beams, 1 MC (b) three MC (c) one MC and two SCS.}}
    \label{fig:realign_all}
\end{figure}

\noindent \textbf{Large number of candidate beams:} In the first scenario, we increase the number of candidate beams to be explored to 81 at the transmit and receive ULA, resulting in overlapping beams. The rest of the parameters are kept identical as in Fig.~\ref{fig:realign}(a).  We observe in Fig.~\ref{fig:realign_all}(a), that the cumulative BER has slightly increased as compared to Fig.~\ref{fig:realign}(a) since the number of beams to be explored has increased here resulting in a higher regret for all algorithms. However, $\mathbf{UCB}_{ISAC}$ still significantly outperforms $\mathbf{UCB}_{SNR}$ and LUCB algorithms.   \\

\noindent \textbf{Large number of multipath components:} Next, we discuss the scenario with greater clutter by considering three MCs with only a single MU and no SCS. Due to the high multipath, the number of candidate beams in this scenario is quite high. As a result, we observe in Fig.~\ref{fig:realign_all}(b) that the cumulative BER is high for each algorithm. Still the performance of $\mathbf{UCB}_{ISAC}$ is better than $\mathbf{UCB}_{SNR}$ and LUCB. This analysis showcases the effectiveness $\mathbf{UCB}_{ISAC}$ in quasi-stationary scenarios in determining the best beam in real-time conditions due to the implementation of the proposed change detection method.  \\

\noindent \textbf{Presence of static clutter scatterers:} Further, we analyze the performance for a scenario where we have two SCS along with one MU and one MC. Note that SCS does not contribute towards communication with the MU and only increases the number of candidate beams. Hence, we observe that there is a significant increase in the cumulative BER of $\mathbf{UCB}_{SNR}$ and LUCB algorithms resulting in poor performance of communication links. However, $\mathbf{UCB}_{ISAC}$ eliminates the beams which correspond to SCS due to their zero Doppler shift, and hence the remarkable difference is observed between the $\mathbf{UCB}_{ISAC}$ and $\mathbf{UCB}_{SNR}$ in Fig.~\ref{fig:realign_all}(c).

\section{Hardware Software Co-design and Fixed-point Analysis}
\label{sec:hardware}

\subsection{Performance and Complexity Analysis}
{\color{black}The mapping of algorithms on edge platforms (for example, on wireless nodes mounted on road infrastructure, unmanned aerial vehicles, etc.) is essential to validate their feasibility on the different types of platforms such as processors and FPGA. The design of fixed-point hardware IPs is the first essential step towards commercialization.} In this section, we design and implement the MAB and RSP algorithms on SoC via hardware-software co-design (HSCD) and fixed-point analysis. {\color{black} The hardware implementation of bandit learning with ISAC provides the estimate of real-time analysis, complexity analysis, and the resource utilization of the ISAC-MAB framework.}
We use the Zynq Ultrascale+ multi-processor SoC (ZMPSoC), ZCU111, comprising of quad-core ARM Cortex A53 processor and Ultrascale+ FPGA from AMD-Xilinx. We compare the resource utilization, power consumption in Watts (W), and execution time in milliseconds (ms) of various architectures of these algorithms obtained via HSCD and fixed-point analysis.

In Table~\ref{tab:MAB_COMP}, we consider 12 different architectures of the MAB algorithms. We consider $\mathbf{UCB}_{SNR}$ with 32 beams and $\mathbf{UCB}_{ISAC}$ with 16 and 8 beams. In Rows 1-3, we compare the execution time of the algorithms on the A53 processor without using any FPGA resources. It can be observed that proposed $\mathbf{UCB}_{ISAC}$ offers a 2-5 factor improvement in execution time over $\mathbf{UCB}_{SNR}$. However, the execution time of all three architectures is significantly large for the selected time slot of 4 ms.  Hence, the acceleration of the MAB algorithm on the FPGA is desired. In Rows 4-6, we implement three algorithms on SoC via HSCD where calculation of UCB quality factor and identification of beam with highest quality factor is realized on the FPGA. Such architecture results in a significant decrease in the execution time for all three algorithms. Note that we have selected low-area architecture where the UCB quality factor of all beams is calculated sequentially using a single hardware block. Hence, the improvement in execution time is achieved due to faster calculation of the UCB quality factor in equation \eqref{equ:UCB} on the FPGA compared to the processor. To further reduce the execution time, allowing more time for data communication, we explored serial-parallel architectures using four parallel hardware blocks to calculate the UCB quality factors of four beams in parallel. For instance, in the case of $\mathbf{UCB}_{SNR}$ with 32 beams, all UCB quality factor calculations will be completed in 8 iterations compared to 32 iterations in the case of Row 4 architecture. In Rows 1-9 architectures, we have used single-precision floating-point (SPFL) word-length (WL). Such high precision may not be needed in many practical applications. In Rows 10-12, we have reduced the word length (WL) to 24 bits, which offered nearly identical regret performance as that of the SPFL architecture but offered further savings in resource utilization.  In the future, we plan to optimize the architecture further to reduce the WL below 18, which can offer substantial savings in resources, especially DSP units that accept two inputs of 18 and 24 bits, respectively.

\begin{table*}[!ht]
\centering
\caption{\footnotesize Resource Utilization, Power Consumption and Execution Time Comparison of MAB Algorithms on SoC}
\label{tab:MAB_COMP}
\renewcommand{\arraystretch}{1.5}
\resizebox{2\columnwidth}{!}{%
\begin{tabular}{|c|c|c|ccccc|c|}
\hline
\textbf{\begin{tabular}[c]{@{}c@{}}Sr.\\ No.\end{tabular}}  & \textbf{Platform} & \textbf{Algorithm} & \multicolumn{1}{c|}{\textbf{LUT}} & \multicolumn{1}{c|}{\textbf{FF}} & \multicolumn{1}{c|}{\textbf{DSP}} & \multicolumn{1}{c|}{\textbf{BRAM}} & \textbf{\begin{tabular}[c]{@{}c@{}}Dynamic\\ Power (W)\end{tabular}} & \textbf{\begin{tabular}[c]{@{}c@{}}Execution Time\\ (ms)\end{tabular}} \\ \hline
1&{\textbf{A53}} & $\mathbf{UCB}_{SNR}$ ($K=32$) & \multicolumn{4}{c|}{{NA}} & 1.5& 15 \\ \cline{1-1} \cline{3-3} \cline{9-9} 
2&\textbf{(SPFL)} & $\mathbf{UCB}_{ISAC}$ ($\tilde{K}=16$) & \multicolumn{4}{c|}{}& & 7.8 \\ \cline{1-1} \cline{3-3} \cline{9-9}
  3& & $\mathbf{UCB}_{ISAC}$ ($\tilde{K}=8$) & \multicolumn{4}{c|}{}& & 3 \\ \hline
4&{\textbf{\begin{tabular}[c]{@{}c@{}}A53 and FPGA \end{tabular}}} & $\mathbf{UCB}_{SNR}$ ($K=32$) & \multicolumn{1}{c|}{{6874}} & \multicolumn{1}{c|}{{6265}} & \multicolumn{1}{c|}{{13}} & \multicolumn{1}{c|}{{0}} & {1.7} & 5.3 \\ \cline{1-1} \cline{3-3} \cline{9-9}
5&\textbf{(Serial)} & $\mathbf{UCB}_{ISAC}$ ($\tilde{K}=16$) & \multicolumn{1}{c|}{} & \multicolumn{1}{c|}{} & \multicolumn{1}{c|}{} & \multicolumn{1}{c|}{} &  & 1.8 \\ \cline{1-1} \cline{3-3} \cline{9-9} 
 6&\textbf{(SPFL)} &$\mathbf{UCB}_{ISAC}$ ($\tilde{K}=8$) & \multicolumn{1}{c|}{} & \multicolumn{1}{c|}{} & \multicolumn{1}{c|}{} & \multicolumn{1}{c|}{} &  & 1.7 \\ \hline
7&{\textbf{\begin{tabular}[c]{@{}c@{}}A53 and FPGA \end{tabular}}} & $\mathbf{UCB}_{SNR}$ ($K=32$) & \multicolumn{1}{c|}{{15492}} & \multicolumn{1}{c|}{{18347}} & \multicolumn{1}{c|}{{52}} & \multicolumn{1}{c|}{{2}} & {2.2} & 1.5 \\ \cline{1-1} \cline{3-3} \cline{9-9} 
8&\textbf{(Serial - Parallel)} & $\mathbf{UCB}_{ISAC}$ ($\tilde{K}=16$) & \multicolumn{1}{c|}{} & \multicolumn{1}{c|}{} & \multicolumn{1}{c|}{} & \multicolumn{1}{c|}{} &  & 0.6 \\ \cline{1-1} \cline{3-3} \cline{9-9}
 9&\textbf{(SPFL)} & $\mathbf{UCB}_{ISAC}$ ($\tilde{K}=8$) & \multicolumn{1}{c|}{} & \multicolumn{1}{c|}{} & \multicolumn{1}{c|}{} & \multicolumn{1}{c|}{} &  & 0.55 \\ \hline
10&{\textbf{\begin{tabular}[c]{@{}c@{}}A53 and FPGA \end{tabular}}} & $\mathbf{UCB}_{SNR}$ ($K=32$) & \multicolumn{1}{c|}{{13649}} & \multicolumn{1}{c|}{{16330}} & \multicolumn{1}{c|}{{44}} & \multicolumn{1}{c|}{{2}} & {2.1} & 1.5 \\ \cline{1-1} \cline{3-3} \cline{9-9}
11&\textbf{(Serial - Parallel)} & $\mathbf{UCB}_{ISAC}$ ($\tilde{K}=16$) & \multicolumn{1}{c|}{} & \multicolumn{1}{c|}{} & \multicolumn{1}{c|}{} & \multicolumn{1}{c|}{} &  & 0.6 \\ \cline{1-1} \cline{3-3} \cline{9-9}
12&\textbf{(24 bit)} & $\mathbf{UCB}_{ISAC}$ ($\tilde{K}=8$) & \multicolumn{1}{c|}{} & \multicolumn{1}{c|}{} & \multicolumn{1}{c|}{} & \multicolumn{1}{c|}{} &  & 0.55 \\ \hline
\end{tabular}%
}
\end{table*}
In Table~\ref{tab:RSP_COMP}, we compare the five different architectures of the RSP algorithm comprising MF for range and MUSIC for Doppler velocity estimation. In Row 1, we have realized complete RSP on the A53 processor with SPFL WL, and as expected, the execution time of 200 ms is significantly large, making ISAC infeasible in practice. However, via HSCD, we have moved the matched filtering (MF)-based range detection to FPGA, resulting in a significant improvement in execution time, as shown in Row 2.  A similar trend is observed when we have moved MUSIC-based Doppler estimation to FPGA instead of MF, as shown in Row 3. In Row 4, we have moved MF and MUSIC to FPGA, thereby achieving an execution time of 3.5 ms. In Row 5, we have reduced the WL to reduce further resource utilization, power consumption, and execution time. Depending on the application requirements, we can further optimize the RSP using fewer packets, fast time samples, fewer Doppler elements in MUSIC, or replacing MUSIC with low complexity Doppler estimation algorithms such as pulse-pair processing. At the architecture level, various approaches such as increasing the FPGA clock from 100 MHz to the maximum possible 240 MHz and optimizing memory access, can lead to further reduction in execution time. 

\begin{table*}[!ht]
\centering
\caption{\footnotesize Resource Utilization, Power Consumption and Execution Time Comparison of RSP Algorithm on SoC}
\label{tab:RSP_COMP}
\renewcommand{\arraystretch}{1.5}
\resizebox{2\columnwidth}{!}{%
\begin{tabular}{|l|cc|c|cccc|c|c|}
\hline
{\textbf{\begin{tabular}[c]{@{}l@{}}Sr.\\ No.\end{tabular}}} & \multicolumn{2}{c|}{\textbf{Platform}} & {\textbf{WL}} & \multicolumn{4}{c|}{\textbf{Resource Utilization}} & {\textbf{\begin{tabular}[c]{@{}c@{}}Dynamic Power\\ (W)\end{tabular}}} & {\textbf{Execution Time (ms)}} \\ \cline{2-3} \cline{5-8}
 & \multicolumn{1}{c|}{\textbf{A53}} & \textbf{FPGA} &  & \multicolumn{1}{c|}{\textbf{LUT}} & \multicolumn{1}{c|}{\textbf{FF}} & \multicolumn{1}{c|}{\textbf{DSP}} & \textbf{BRAM} &  &  \\ \hline
1 & \multicolumn{1}{c|}{MF,MUSIC} & - & {SPFL} & \multicolumn{4}{c|}{NA} & 2.55 & 200 \\ \cline{1-3} \cline{5-10} 
2 & \multicolumn{1}{c|}{MUSIC} & MF &  & \multicolumn{1}{c|}{22024} & \multicolumn{1}{c|}{22024} & \multicolumn{1}{c|}{78} & 260 & 3.146 & 22 \\ \cline{1-3} \cline{5-10} 
3 & \multicolumn{1}{c|}{MF} & MUSIC &  & \multicolumn{1}{c|}{32450} & \multicolumn{1}{c|}{32450} & \multicolumn{1}{c|}{316} & 14.5 & 3.192 & 15 \\ \cline{1-3} \cline{5-10} 
4 & \multicolumn{1}{c|}{-} & MF,MUSIC &  & \multicolumn{1}{c|}{51182} & \multicolumn{1}{c|}{51182} & \multicolumn{1}{c|}{394} & 274.5 & 3.454 & 3.5 \\ \hline
5 & \multicolumn{1}{c|}{-} & MF,MUSIC & \begin{tabular}[c]{@{}c@{}}FFT: 24 bit\\ IFFT: 16 bit\\ MUSIC: SPFL\end{tabular} & \multicolumn{1}{c|}{\begin{tabular}[c]{@{}c@{}}45333\\ (-11.4\%)\end{tabular}} & \multicolumn{1}{c|}{\begin{tabular}[c]{@{}c@{}}45333\\ (-11.4\%)\end{tabular}} & \multicolumn{1}{c|}{\begin{tabular}[c]{@{}c@{}}370\\ (-6\%)\end{tabular}} & \begin{tabular}[c]{@{}c@{}}122\\ (-55.6\%)\end{tabular} & \begin{tabular}[c]{@{}c@{}}3.294\\ (-4.6\%)\end{tabular} & 3.3 \\ \hline
\end{tabular}%
}
\end{table*}
\subsection{Timing analysis} 
Using the execution time results obtained from hardware implementation, we compare the exploration time of two algorithms and revisit the throughput calculation discussed in Table~\ref{tab:throughput_point} and Table~\ref{tab:throughput_quasi}. We consider $K=32$ candidate beams spanning from -64$^o$ to 64$^o$, and the total time slots considered here are 2000, corresponding to a total duration of 8 seconds. We consider a scenario with one MU and one MC.  

Firstly, we discuss the exploration time in the $\mathbf{UCB}_{SNR}$ algorithm, where the duration of each time slot is fixed at 5.5 $ms$. Here, 1.5 $ms$ corresponds to beam selection using $\mathbf{UCB}_{SNR}$ in each time slot as shown in Row 10 of Table~\ref{tab:MAB_COMP}, and the remaining 4 $ms$ duration includes the generation and transmission of the DL packet at BS, receiver processing time at MU and the feedback over UL packet from MU. Since each beam is explored once over a time slot, the total exploration time of $\mathbf{UCB}_{SNR}$ is 128 $ms$. Next, we discuss the exploration time in the $\mathbf{UCB}_{ISAC}$ algorithm. In the RSP phase, each radar packet is transmitted by the BS with a 50$\%$ duty cycle, resulting in a $T_p$ of 0.58 $\mu$s. We consider one CPI of 11.6 $\mu$s, which comprises 20 radar packets and 32 sub-slots (corresponding to 32 beams) each of the duration of CPI results in a total time of 0.4 $ms$. This is significantly smaller than 3.5 $ms$ required for the RSP of each beam, as shown in Row 4 of Table~\ref{tab:RSP_COMP}. In 3.5 $ms$, 2 $ms$ are required for MF and 1.5 $ms$ for MUSIC. Since the radar-reflected signals of all beams are available, we can perform the RSP in a pipelined fashion, which means the RSP of subsequent beams can be initiated at an interval of 2 $ms$. Thus, the total RSP time is at most 67 $ms$. With an average of eight ($\tilde{K}=8$) beams, we can start the exploration at the 54 $ms$, and hence, the total exploration time of the  $\mathbf{UCB}_{ISAC}$ is 84 $ms$. We observe from this exercise that the reduction in exploration time in $\mathbf{UCB}_{ISAC}$ is approximately 35\% of the exploration time in $\mathbf{UCB}_{SNR}$.  With more efficient and faster architecture, we can potentially achieve a 65\% reduction in exploration time, and further reduction can be made by using multiple RSP architectures in parallel. 

The use of RSP also allows more time for the data communication by allowing the UCB to focus on $\tilde{K}$ instead of $K$ beams. Out of the total 2000 time slots, the minimum time slot duration for UCB is 5.5 $ms$ since 4 $ms$ are required for data communication and 1.5 $ms$ for UCB in each time slot. Since UCB is not in the first 32 time slots, we can transmit the data over 5.5 $ms$  and over 4 $ms$ in the rest of the 1968 time slots. In $\mathbf{UCB}_{ISAC}$, the first slot is used for the radar waveform communication and thereafter 12 slots are unused since the RSP is being carried out. Thereafter, 8 slots are used for exploring $\tilde{K}=8$ beams with a time slot duration of 5.5 $ms$ and in the remaining 1980 slots where UCB is used for 0.55 $ms$ followed by data communication over the remaining 4.95 $ms$. As shown in Table~\ref{tab:throughput_point_hardware}, this results in substantial improvement by factor 1.4 in throughput over $\mathbf{UCB}_{SNR}$. The gain in improvement increases with the increase in transmission bandwidth.

\begin{table}[htbp]
\centering
\caption{\footnotesize Comparison of Throughput of different algorithms and targets models in stationary case based on hardware analysis }
\begin{tabular}{|l|lll|}
\hline
{\textbf{Algorithms}} & \multicolumn{3}{c|}{\textbf{Throughput (Mbps)}}                     \\ \cline{2-4} 
                                     & \multicolumn{1}{l|}{\textbf{Single point}} & \multicolumn{1}{l|}{\textbf{Pedestrian}} & \textbf{Mid-size car} \\ \hline 
                                     $\mathbf{UCB}_{ISAC}$               & \multicolumn{1}{l|}{8.59}                  &  \multicolumn{1}{l|}{8.41}   &   7.99                   \\ \hline

$\mathbf{UCB}_{SNR}$                & \multicolumn{1}{l|}{6.02}                  & \multicolumn{1}{l|}{6.02}   &   6.02                   \\ \hline
\end{tabular}
\label{tab:throughput_point_hardware}
\end{table}

\section{Conclusion}
\label{sec:Conclusion}
In this paper, we demonstrate how MAB enhanced with ISAC can improve the overall communication metrics. Since the radar and communication share spectrum, hardware, and waveform; the communication MU are among the detected mobile radar targets. Thus instead of searching all the candidate beams for MAB, we propose that only those beams that indicate the presence of mobile radar targets are scanned.  
Simulation as well as experimental results show improved BER and throughput for the ISAC-MAB algorithm as compared to the conventional MAB.  In our work, we limit our discussion to the upper confidence bound-based algorithm and provide regret bounds. The proposed idea can be easily extended to other regret minimization-based MAB algorithms, such as UCB variants, Kullback–Leibler (KL) UCB, and Thompson Sampling as well as pure exploration-based MAB algorithms, such as lower upper confidence bound (LUCB) algorithm. {\color{black} Moreover,  MmW communication is often affected by dynamic and adversarial conditions, such as changing channel conditions, obstacles, and interference.  Adversarial bandit models can be explored in future work for the ISAC-MAB framework to adapt to these conditions by treating them as adversaries and designing algorithms robust to such variations.}
\bibliographystyle{ieeetran}
\bibliography{reference}    

\end{document}